\documentclass[preprint2]{aastex}

\shorttitle{Accelerated Black Holes}
\shortauthors{Kornreich \& Lovelace}

\newcommand{\etal}{ {\it et al.} }
\begin{document}

\title{Dynamics of Kicked and Accelerated Massive Black Holes in Galaxies}

\author{David A. Kornreich\altaffilmark{1}}
\affil{Department of Physics and Astronomy, Humboldt State University, Arcata, CA 95521}
\and
\author{Richard V. E. Lovelace}
\affil{Departments of Astronomy and Applied and Engineering Physics, Cornell University, Ithaca, NY 14850}

\altaffiltext{1}{Visiting Astronomer, Arecibo Observatory. The Arecibo
Observatory is part of the National Astronomy and Ionosphere Center,
which is operated by Cornell University under a cooperative agreement
with the National Science Foundation}

\begin{abstract}

       A study is made of the behavior of massive black holes in disk
galaxies that have received an impulsive kick from a merger or a
sustained acceleration from an asymmetric jet.  The motion of the gas,
stars, dark matter, and massive black hole are calculated using the
GADGET-2 simulation code.  The massive black hole escapes the galaxy
for kick velocities above about $600$~km~s$^{-1}$ or accelerations
above about $4\times 10^{-8}$~cm~s$^{-2}$ over time-scales of the
order of $10^8$ yr.  For smaller velocity kicks or smaller
accelerations, the black hole oscillates about the center of mass with
a frequency which decreases as the kick velocity or acceleration
increases.  The black hole displacements may give rise to observable
nonaxisymmetries in the morphology and dynamics of the stellar and
gaseous disk of the galaxy.  In some cases the dynamical center of the
galaxy is seen to be displaced towards the direction of the BH
acceleration with a characteristic ``tongue--'' shaped extension of
the velocity contours on the side of the galaxy opposite the
acceleration.

\end{abstract}


\section{Introduction}

   Recent breakthroughs in numerical General Relativity have led to
predictions of large recoil velocities of merged binary black-holes
(BHs) as the binary radiates away linear momentum as gravitational
waves during the final stages of merger
(\citet{G07},\citet{Campanelli}).  
The velocity or ``kick'' of the merged black-hole in
the nucleus of a galaxy can be as large as $4000$~km~s$^{-1}$, which
would be more than enough
to eject the BH from the galaxy.  However, it is more likely that
typical kick velocities are much smaller ($\lesssim 200$~km~s$^{-1}$)
owing to gas accretion by the merging BHs and a large ratio of the
masses of the two initial BHs \citep{Bogdanovic}.

In another situation, the asymmetric or one-sided jet from the disk
around a massive BH can give a steady acceleration of the black hole
over the Salpeter time-scale $\sim 10^8$ yr (\citet{Shklovsky};
\citet{Tsygan07}) and this may push the BH out of the galaxy.
\citet{Wang92} pointed out that some combinations of dipole and
quadrupole symmetry magnetic field components threading the accretion
disk of the BH can give a much stronger magnetically driven jet on one
side of the disk than on the other.  It is of interest that a bright
quasar has been discovered which is {\it not} inside a massive galaxy
\citep{Magain} and so it may have been ejected from a galaxy.

Field galaxies themselves may be morphologically or dynamically
nonaxisymmetric even in the absence of tidal companions
\citep{Baldwin}. Up to 30\% of field galaxies are nonaxisymmetric
(\citet{K98}, \citet{ZR97}). If central supermassive BHs receive impulses from
mergers or asymmetric jets, causing motion in the host galaxy, some of
that motion may be transmitted to the galaxy via dynamical
friction. We may expect that in such galaxies, morphological or
dynamical effects may be seen in the stellar or gas distribution as a
result of this transfer of momentum.

The objective of this work is to determine the BH displacement as a
function of time for different initial kick velocities and directions,
different accelerations due to a one sided jet, and different BH
masses.  A second objective is to study influence the BH displacement
has on the morphology and the kinematics of the host galaxy and
determine whether nonaxisymmetries in disk galaxies are attributable
to motions of the BH. We have chosen to concentrate on the morphology
of disk galaxies rather than ellipticals because the dynamics of disk
galaxies, being supported by bulk rotation rather than random motions,
are more likely to exhibit signs of disturbance if the motion of the
BH interferes with that rotation.

    Section~\ref{Methods} of the paper describes the initial conditions and the
numerical methods used.  Section~\ref{Kicks} discusses the results, first for
the case of velocity kicks of a $10^8 M_\odot$ BH, and then of a $10^9
M_\odot$ BH.  Section~\ref{BHAccel} discusses the results for cases of one-sided
jets, first for the $10^8 M_\odot$ BH, and then for the $10^9 M_\odot$
BH.  Section~\ref{Results} gives the conclusions of this work.

\section{Numerical Methods and Initial Conditions\label{Methods}}

The computations for this experiment were performed using the N--body,
Smoothed Particle Hydrodynamics (SPH) GADGET--2 code of
\citet{S05}. We have added the ability to assign a particular particle
an arbitrary acceleration, in addition to the calculated gravitational
acceleration, in order to simulate forces due to an asymmetric
jet. This particle is given significant mass and placed initially
at the center of the galaxy.

   The considered galaxy consists of a dark halo, a stellar disk
and bulge, and a gas disk. The gas particles are treated
gravitationally and hydrodynamically, while the other particles
interact gravitationally. The halo, disk, and bulge particles differ
in mass and spatial distribution, but are otherwise treated
identically in the computation. The central stellar bulge is modeled
as a \citet{Plummer} sphere, which has a density distribution of a
relaxed $n=5$ polytrope given by:
\begin{equation}
\rho(r) = \frac{3}{4\pi}\frac{M_B}{R_B^3}\frac{1}{\left[1+\left(r/R_B\right)^2\right]^{5/2}}~,
\end{equation}
with bulge scale radius $R_B = 1$~kpc and mass $M_B = 10^{10}M_\odot$.

This bulge is placed at the center of a softened isothermal dark
matter halo of density profile 
\begin{equation}
\rho(r) = {\rho_0 \over 1 + (r/r_c)^2}~,
\end{equation}
total mass $5.6\times 10^{11}M_\odot$, with a cutoff radius $50$~kpc,
and core radius $r_c=2$~kpc. A stellar and gas disk of
scale radius $r_0=3.5$~kpc and cutoff radius $R=50$~kpc of exponential surface density
\begin{equation}
\Sigma(r) = \frac{M_D}{2\pi r_0^2\left[
    1-e^{-R/r_0}\left(1+R/r_0\right)\right]}e^{-r/r_0}
\end{equation}
surrounds the bulge with total mass $M_D=2.8\times 10^{10}M_\odot$, at
a gas mass fraction of $25\%$. In the vertical direction, the disk
particles are given velocity dispersions and uniformly random initial
positions corresponding to a disk of 1~kpc thickness.

The total dynamical mass of the galaxy model is therefore $M_{\rm tot}
=5.98\times 10^{11}M_\odot$. This model was chosen to conform to
parameters observed in giant disk galaxies such as those described by
\citet{Rownd} and \citet{Dickey}, in particular, a circular velocity
of $220$~km/s with a $10$~km/s velocity dispersion in the disk.  The
total number of simulation particles used in the model was $2\times
10^6$ for most model runs.

In one sequence of runs, we gave the central black hole an initial
impulsive velocity ``kick'' to simulate the recoil from the merging of
two black holes.  In one sequence, the kick was in the plane of the
galaxy, and in the other, it was normal to the galaxy plane. The black
hole was given no further acceleration.  In another sequence of runs,
the BH was given a steady acceleration of order $10^{-8}$~cm~s$^{-1}$
for a time of the order of the Salpeter time in order to simulate the
influence of a one-sided jet.  Each of these sequences was run for a
BH mass of $10^8M_\odot$ and a larger black hole of $10^{9}M_\odot$.

During each run, the position, velocity, and acceleration of the BH
was output for each timestep. Data dumps containing positions,
velocities, accelerations, and (for the gas particles) temperatures
were generated periodically, usually every $0.01$~Gyr of simulation
time. These data were read into the IDL data processing environment,
where it was possible to display the results as seen from a variety of
viewing angles. For a given viewing angle, gas particles were
convolved with a Gaussian simulated telescope beam and displayed as
logarithmic intensity contours overlaid on a diagram of stellar
particle positions. Gas velocities were similarly convolved and used
to produce radial velocity maps.

Additional sample runs were conducted with isothermal halo core radius
0.5~kpc, with no radius, and with the halo density profile of
\citet{NFW}. In each of these cases, the results were similar and the
details of the behavior did not substantially depend on the nature of
the halo density profile.

\section{Velocity ``Kicks''\label{Kicks}}
\subsection{Black Hole Mass $10^8M_\odot$}
In this series of model runs, the BH was placed near the center of
mass of the galaxy and given an initial velocity in the plane of the
disk. The series of initial velocities started at zero and progressed
in increments of $50$~km~s$^{-1}$ up to a velocity of 650~km~s$^{-1}$,
for which the central BH escaped the galaxy. It is interesting to note
that this critical kick velocity is significantly larger than the
escape velocity of the center of the galaxy, which was measured to be
450~km~s$^{-1}$. We attribute this excess to dynamical friction.

\begin{figure}
\epsscale{.7}
\plotone{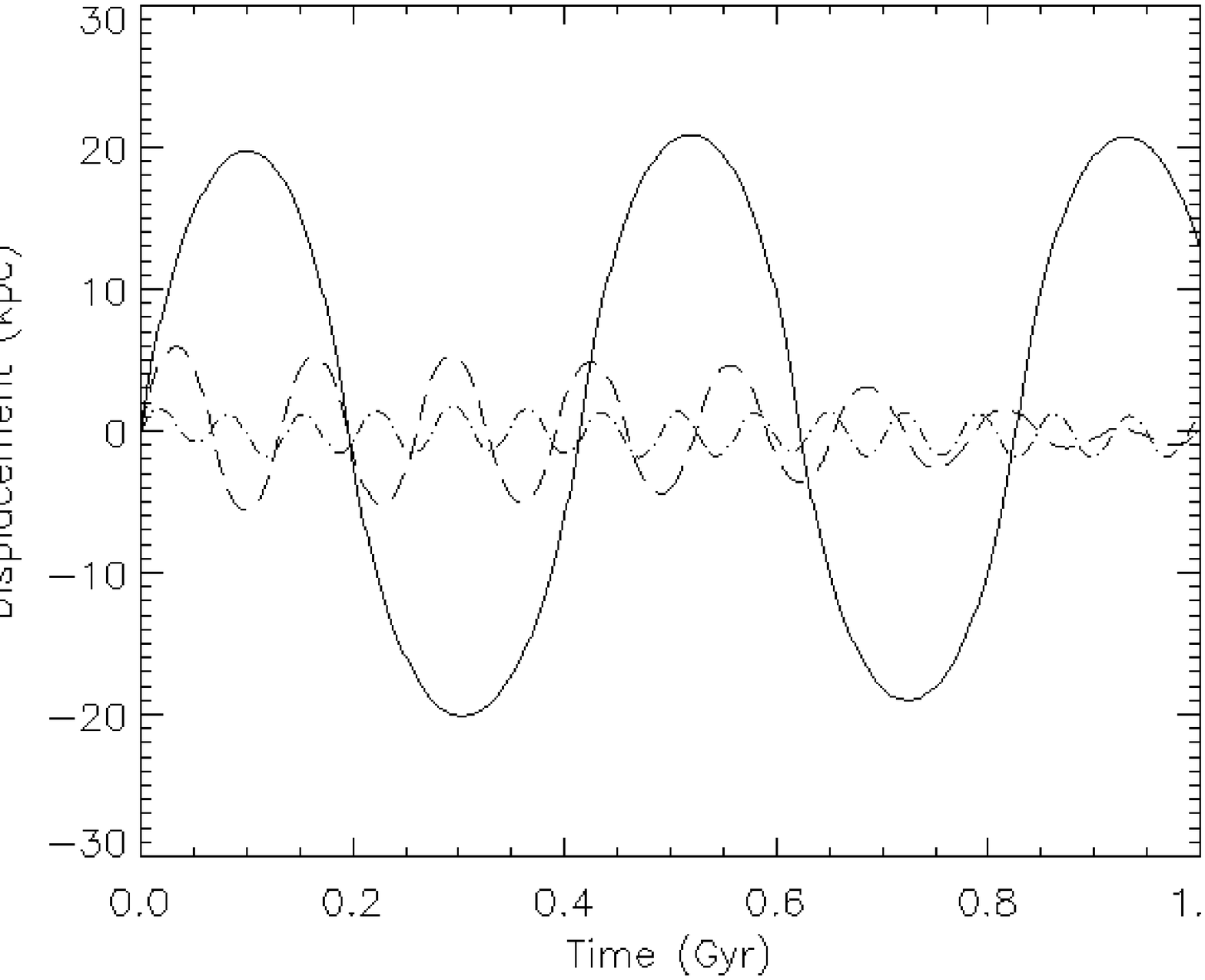}\\
\plotone{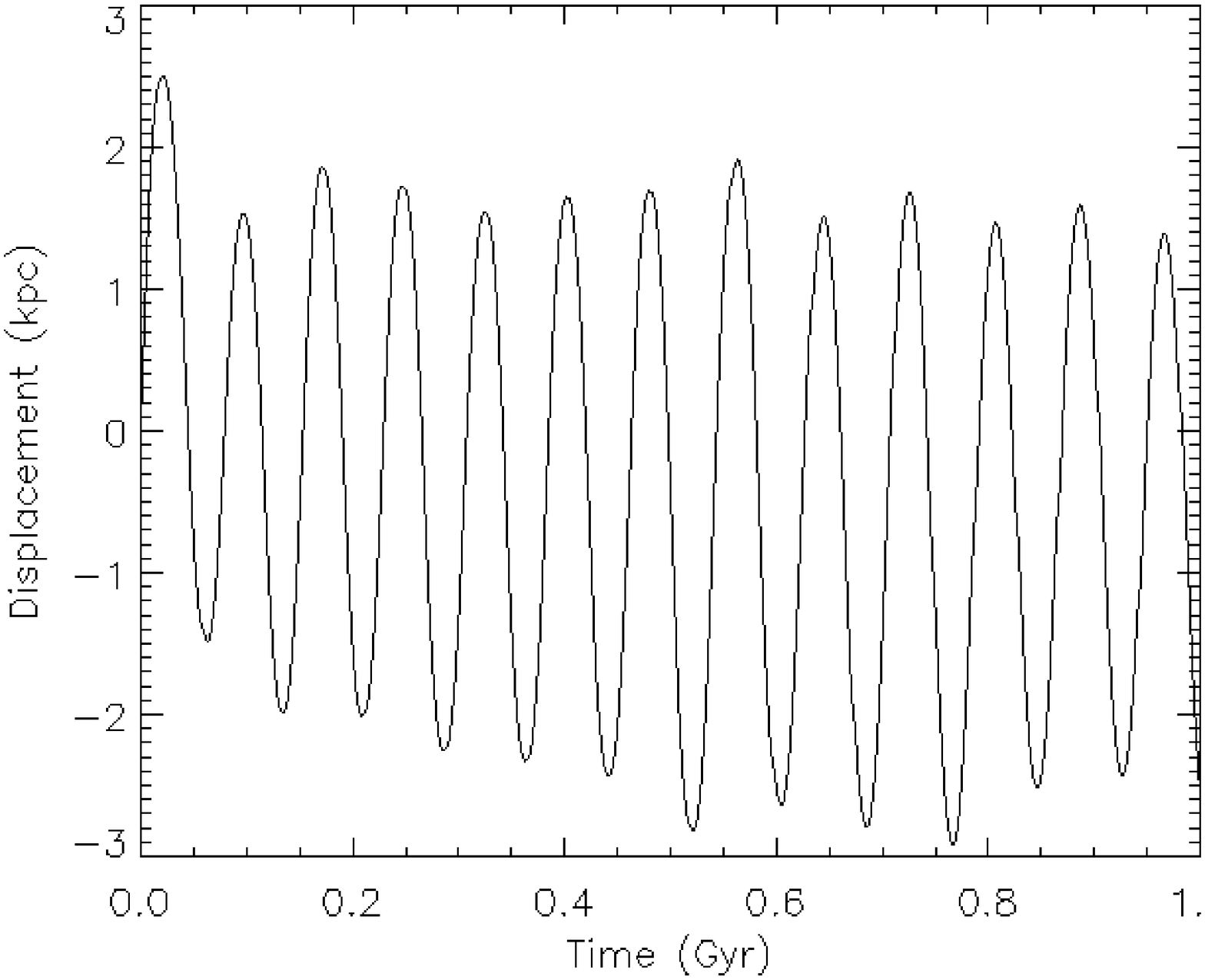}
\caption{The top  panel shows the motion along the 
direction of kick (in the plane of the galaxy)  for a BH of mass
  $10^8M_\odot$  for an initial velocity kick
  of $500$~km~s$^{-1}$ (solid line), $350$~km~s$^{-1}$ (dashed line), and
  $150$~km~s$^{-1}$ (dash--dot line).   The bottom panel is 
for an initial kick velocity $200$~km~s$^{-1}$.
\label{xt}}
\end{figure}

\begin{figure}
\plotone{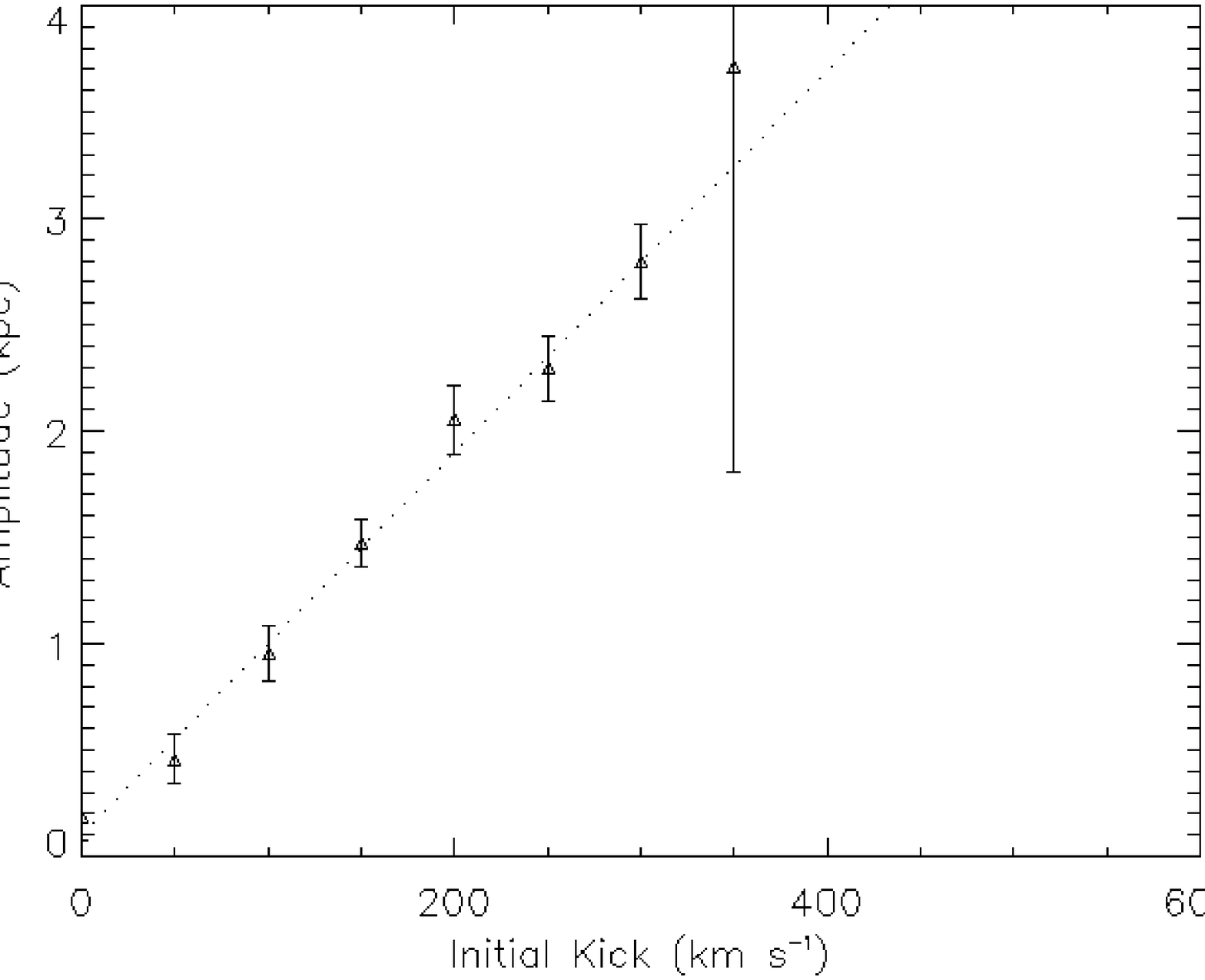}\\
\vskip 1cm
\epsscale{0.6}
\hskip -1cm \plotone{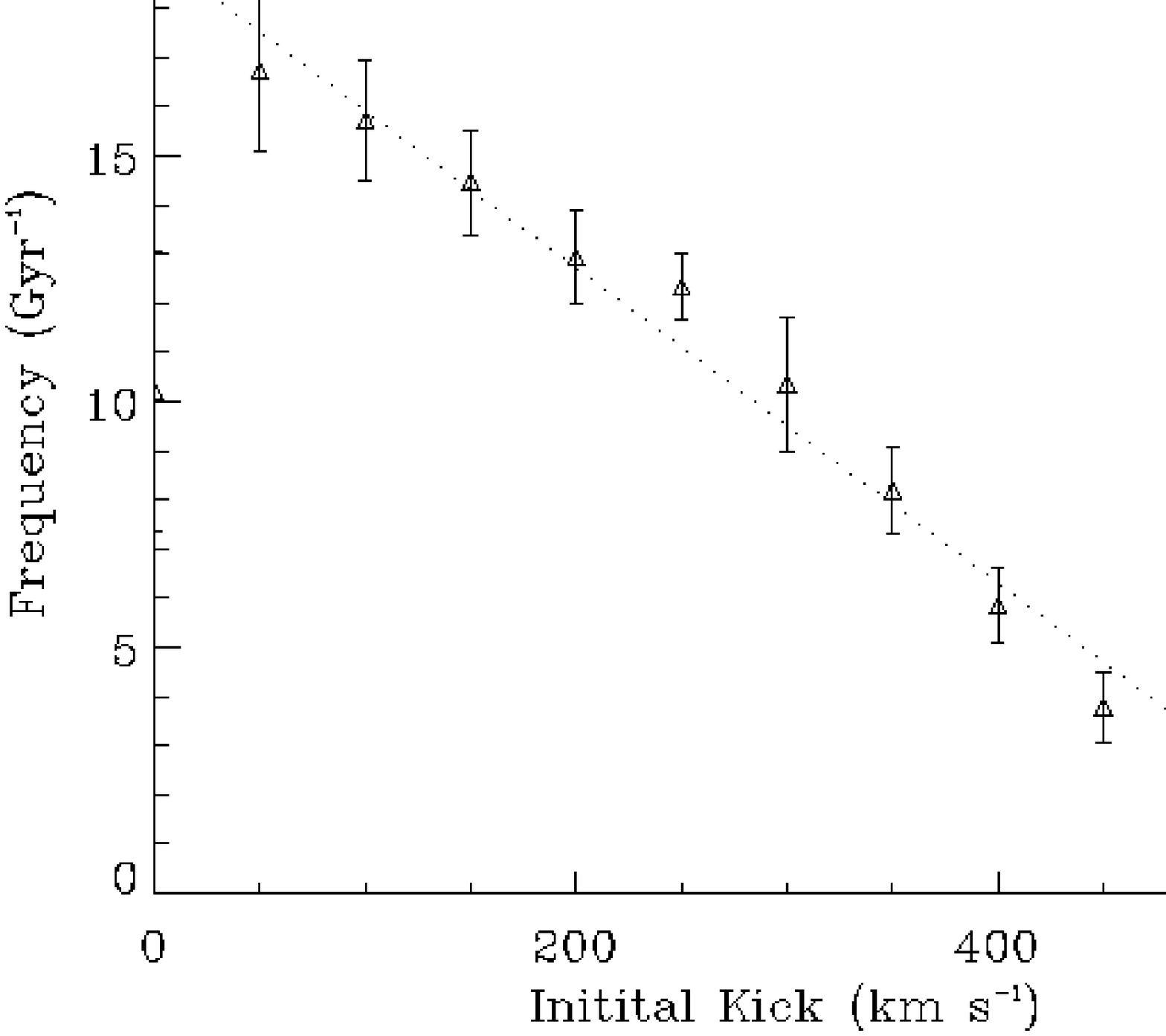}
\caption{Amplitudes and frequencies of the central BH motion
  resulting from velocity kicks in the plane of the galaxy for a $10^8~M_\odot$ BH.}
\label{amplitude}
\end{figure}

The general behavior of the BH after receiving the kick was to
oscillate about the model center of mass approximately sinusoidally
(Figure~\ref{xt}). The amplitude and frequency of the motion remain
roughly constant over a period of $0.5$~Gyr, but depend on the kick
magnitude as shown in Figure~\ref{amplitude}. Error bars represent the
$1\sigma$ standard deviation of these values as calculated for each
cycle. For kick velocities below 400~km~s$^{-1}$, the relation between
amplitude and initial velocity $V_0$ is linear, with $A(V_0)/{\rm kpc}
= (9.30 \pm 0.53)\times 10^{-3} V_0/\mathrm{km\:s^{-1}} + (0.088\pm
0.082)$. For initial velocities above 400~km~s$^{-1}$, the BH
amplitude is beyond the visible disk and becomes exponentially large
until it escapes at 650~km~s$^{-1}$. The frequency of the oscillation
is also linear with $f(V_0)/{\rm Gyr^{-1}} = (17.64 \pm 0.57) -
(0.0317 \pm 0.0015) V_0/\mathrm{km\:s^{-1}}$.

Smaller kicks resulted in the BH remaining in the more massive parts
of the galaxy within 10~kpc for their entire cycles, allowing momentum
from the kick to be efficiently transferred to the galaxy via
dynamical friction. In these cases, damping of the oscillation was on
timescales of $0.7$~Gyr. The galaxy center of mass thus gained
velocity in the direction of kick, resulting in the offsets depicted
in the low kick amplitude curves in Figure~\ref{amplitude}. Larger
kicks, while imparting greater momentum to the BH, caused it to spend
a great deal of time in the thin regions of the galaxy beyond
$r=10$~kpc. This greatly reduces the efficiency of the dynamical
friction, lengthening damping timescales to $2.0$~Gyr and reducing the
momentum transferred to the galaxy.

\begin{figure}
\epsscale{1}
\plotone{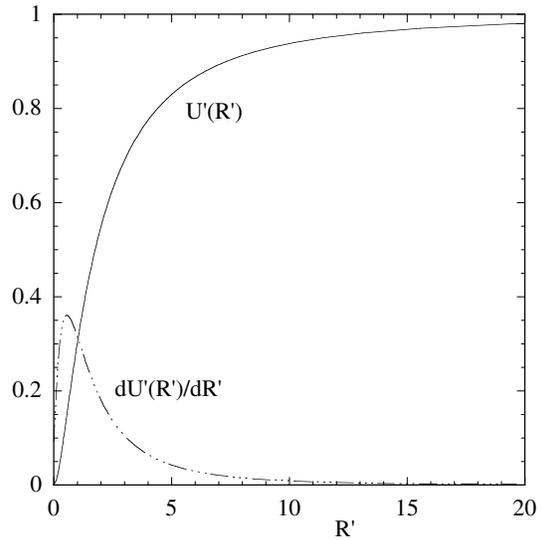}
\caption{Effective potential for the  BH as it
  oscillates about the galaxy center of mass after receiving an
  impulsive kick.  The effective restoring force
  is $\propto -dU_{\rm eff}^\prime/dR^\prime$. 
  The maximum of the force is $\approx 0.36$ at a 
  distance $R^\prime \approx 0.56$.}
\label{frequencies}
\end{figure}

With  the frequency of the BH motion $\omega=2\pi f$ 
known as a function of the amplitude of its motion, 
we can derive the effective potential for its motion, $U_{\rm eff}$.
For a one-dimensional oscillator,
\begin{equation}
R={1\over \sqrt{2}}\int_0^{U_{\rm eff}} { dE \over \omega(E) \sqrt{U_{\rm eff}-E} }~,
\label{RU}
\end{equation}
\citep{LL}.   A first approximation to our simulation
results gives $\omega=\omega_0(1-V_0/V_m)$, where $V_0$ is
the BH kick velocity such that $E=V_0^2/2$, and where $\omega_0$ and $V_m$ are
constants characteristic of the galaxy.   For this dependence,
equation~(\ref{RU}) can be integrated to give
$$
R^\prime={2\over\sqrt{1-U^\prime}} \times 
$$
\begin{equation}
\left[\tan^{-1}\left({1-\sqrt{U^\prime}
  \over \sqrt{1-U^\prime}}\right) + \tan^{-1}\left({\sqrt{U^\prime}
  \over \sqrt{1-U^\prime}}\right) \right]-{\pi \over 2}~,
\end{equation}
where $R^\prime \equiv R/R_0$, $R_0\equiv V_m/\omega_0$ and $U^\prime
\equiv U_{\rm eff}/(V_m^2/2)$. For the case of Figures~\ref{xt}
and~\ref{amplitude}, $V_m\approx 555$~km~s$^{-1}$, $\omega_0\approx
3.51\times 10^{-15}$~s$^{-1}$, and $R_0\approx 5.13\times 10^3$~kpc.
   Of course, for $V_0 \ll V_m$, $U_{\rm eff} =(1/2) \omega_0^2 R^2$ to
a good approximation.
Figure~\ref{frequencies} shows this function and its derivative which
is the negative of the effective force, in contrast to the purely
logarithmic potential of the halo isothermal sphere.

\begin{figure}
\epsscale{.7}
\plotone{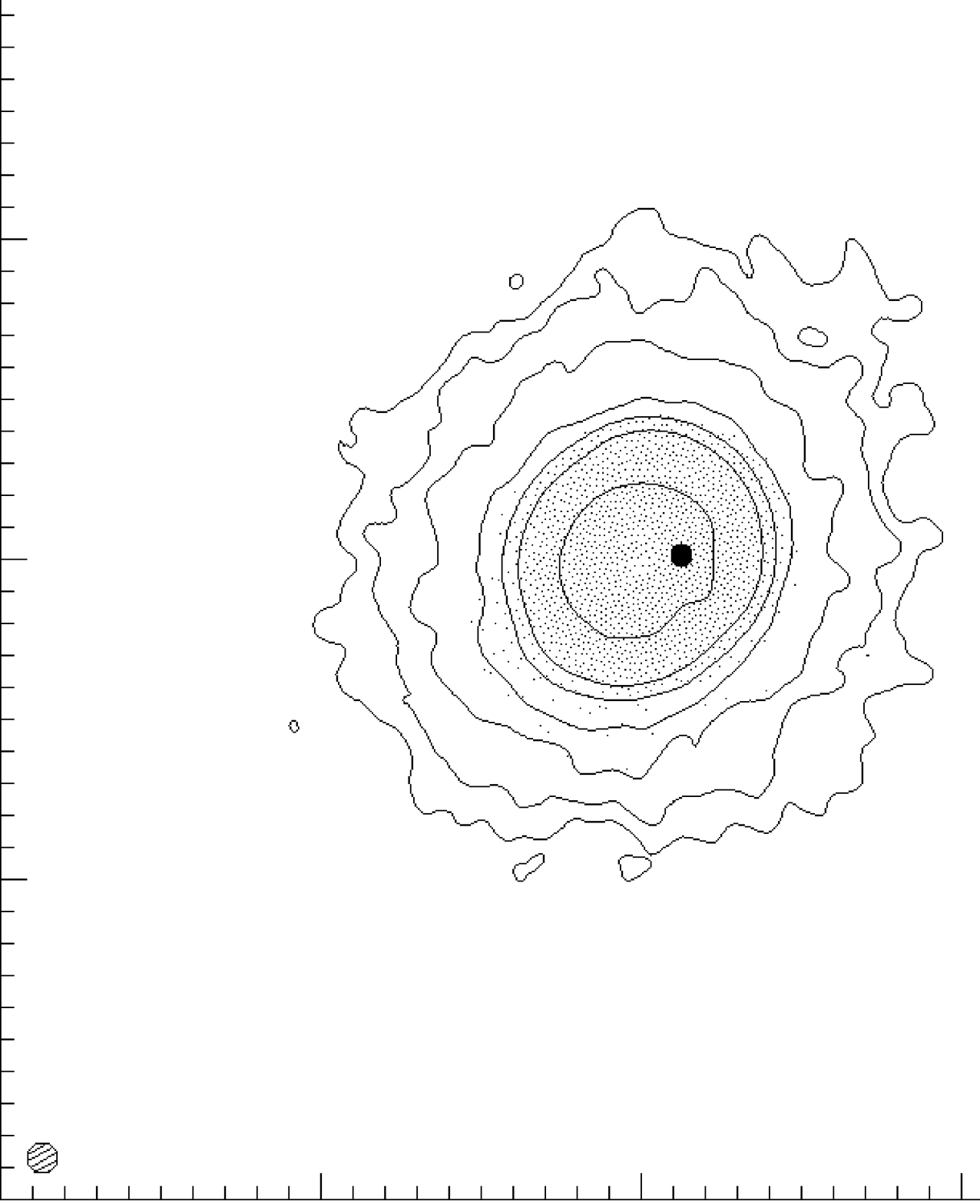}\\
\plotone{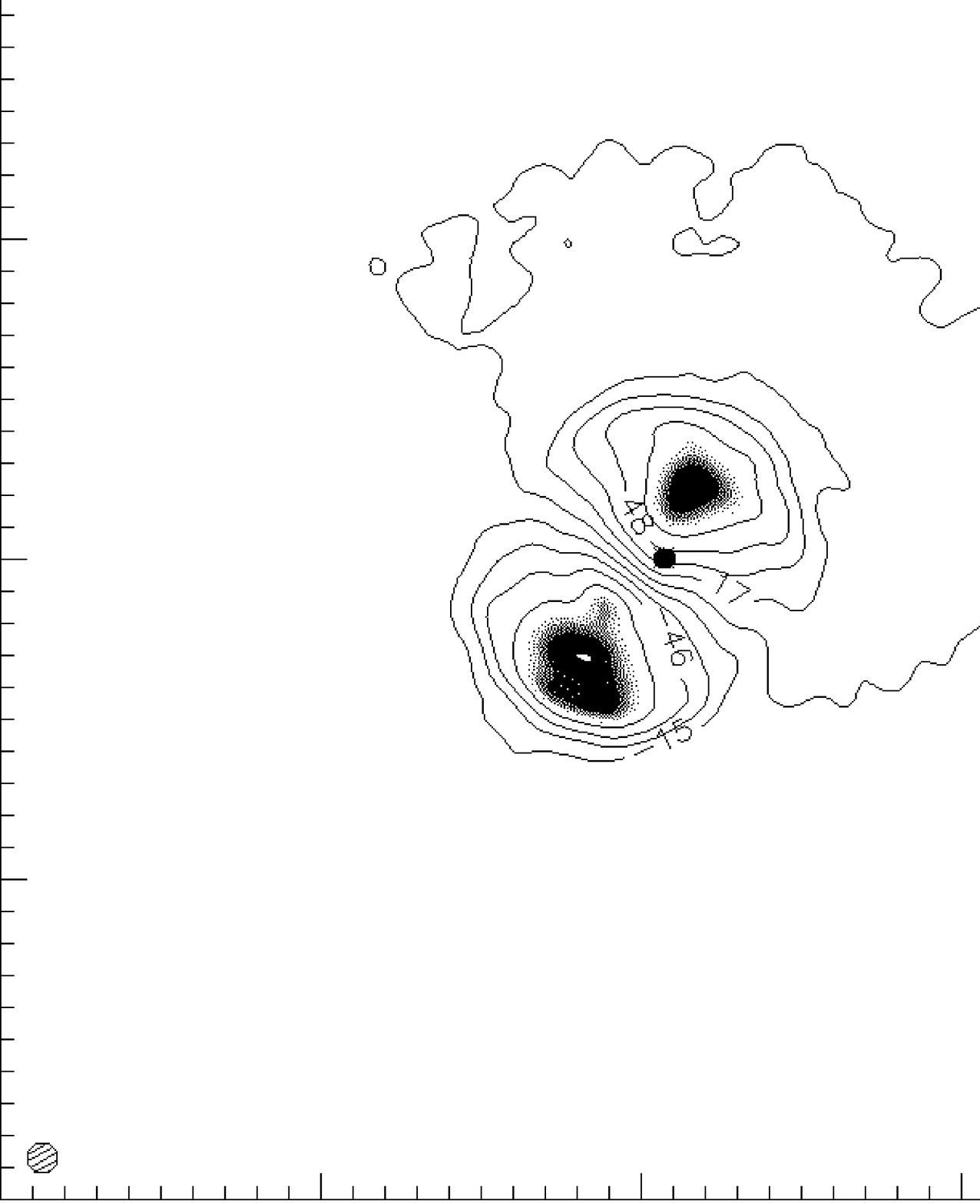}
\caption{Morphology and kinematics of the galaxy model at $t=0.12$~Gyr
  after the central BH of mass $10^8~M_\odot$ has been given a velocity kick of
  $300$~km~s$^{-1}$ in the plane of the galaxy. In the top panel, the
  stellar disk is illustrated in greyscale and the logarithmic contour
  lines indicate density of HI gas, after convolution with a simulated
  Gaussian beam (lower left). 
     In the lower panel, the convolved radial velocity
  of the gas is indicated by the contour lines, labeled in km~s$^{-1}$.     
    The inclination to the line of sight is
  $20$$^\circ$. In both diagrams, the BH is
  indicated by the black dot slightly to the right of center. Major
  tick marks delineate 25~kpc.}
\label{HI}
\end{figure}

We have taken an example frame from the 300~km~s$^{-1}$ model run,
inclined it to the line of sight, and convolved the gas component with
a Gaussian beam. The result is Figure~\ref{HI} which depicts the
morphology and kinematics of the radio gas component of our galaxy
model. Although the central BH is in motion, observable
effects on the host galaxy are minimal. The simulated HI gas is
distributed axisymmetrically, for instance. Moreover, as can be seen
from the figure, lines of constant radial velocity are straight and
parallel through the dynamical center of the galaxy, indicating pure
rotation with little or no streaming motion, even though the central
BH is clearly displaced from the dynamical center of the
galaxy. This is the behavior we observed throughout this set of runs.

\begin{figure}
\epsscale{.8}
\plotone{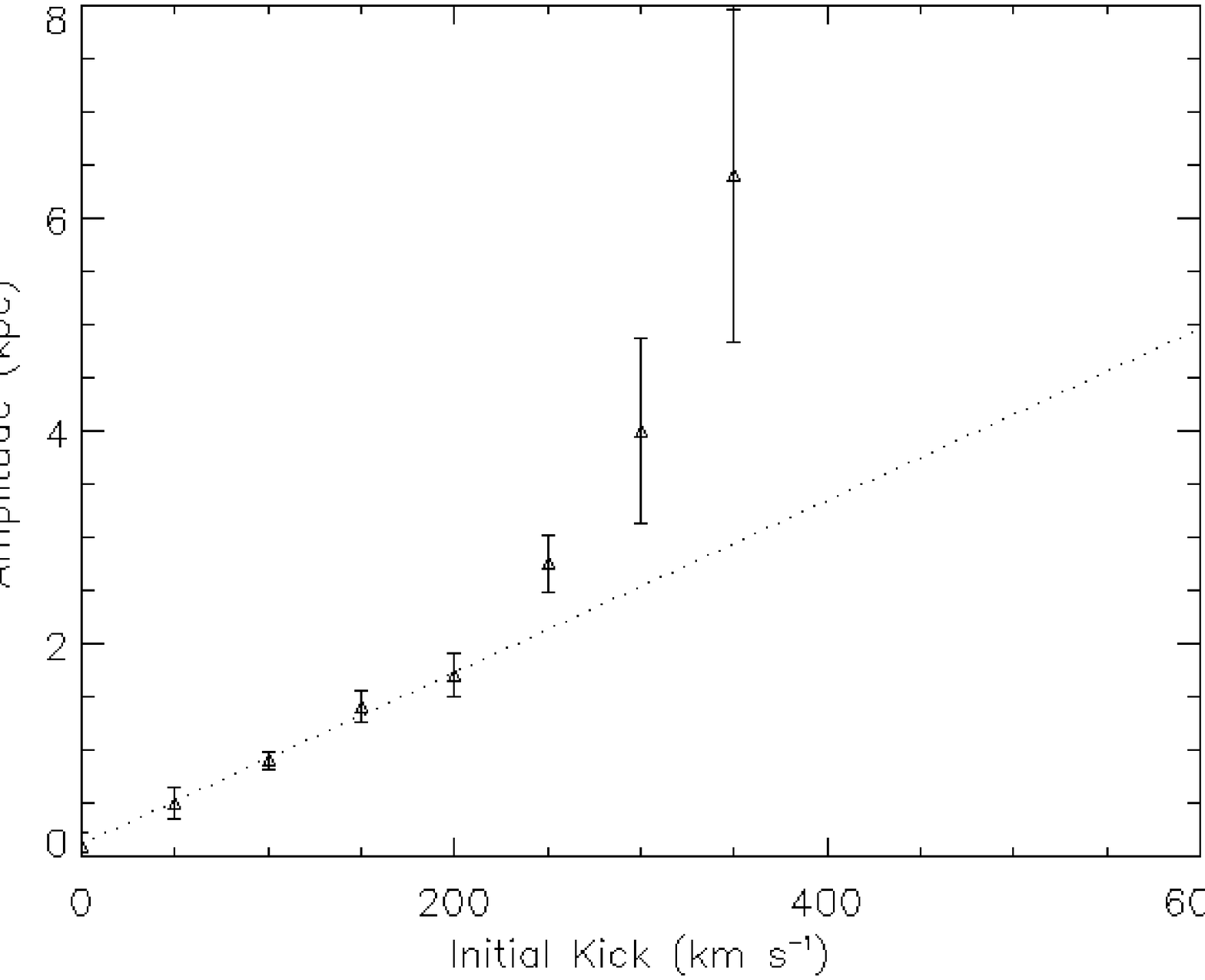}\\
\epsscale{.7}
\vskip 1cm
\hskip -1cm \plotone{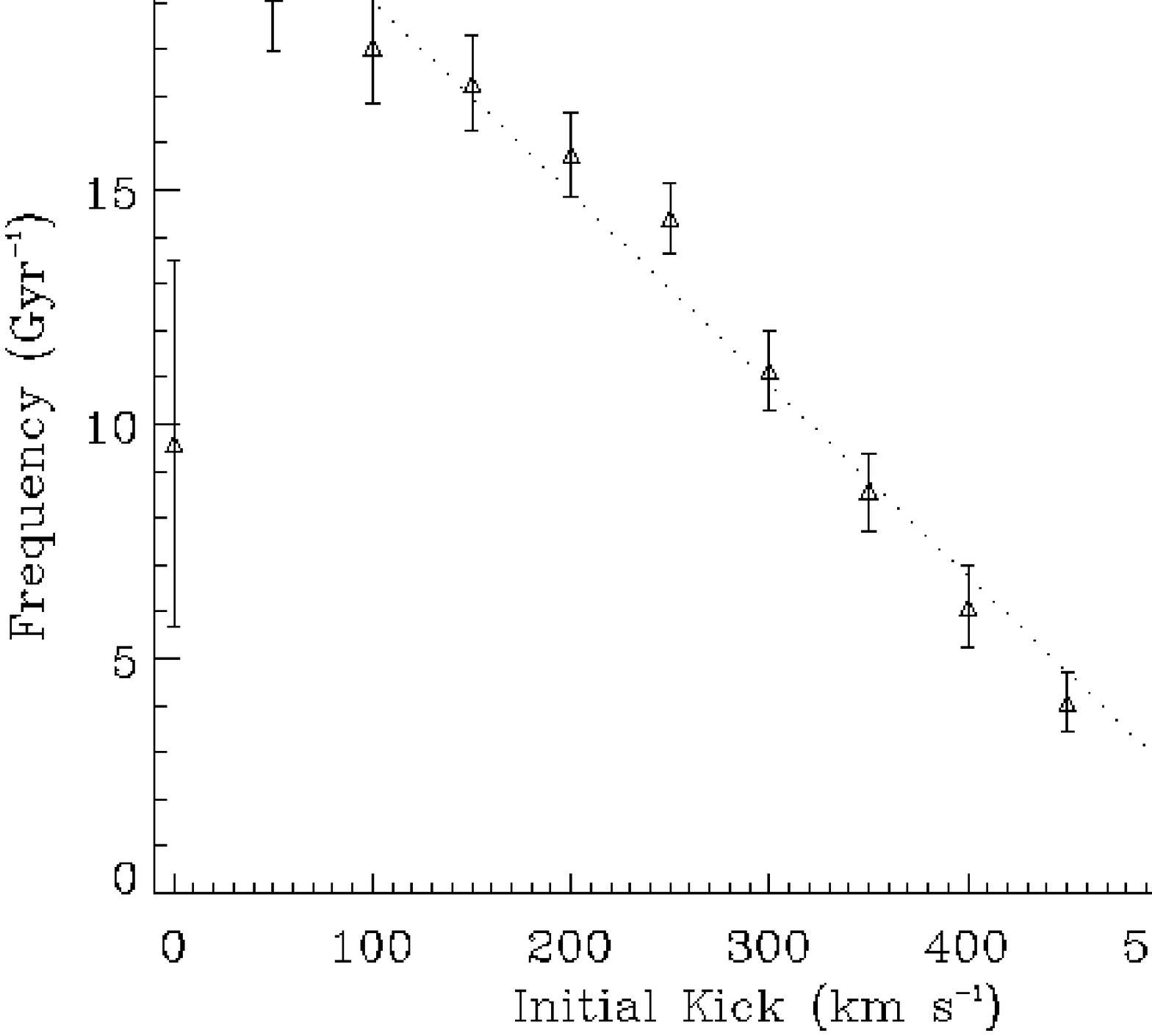}
\caption{Amplitudes and frequencies of the  BH motion
  resulting from velocity kicks perpendicular to the plane of the galaxy for a BH of mass $10^8~M_\odot$.}
\label{amplitude2}
\end{figure}

A series of runs was also conducted with the velocity kick normal to
the plane of the galaxy. Qualitative results were similar to the
case where the kick is in the plane of the galaxy, with the BH
oscillating roughly sinusoidally through the center of mass. The
amplitude--initial kick relation remains linear for a slightly smaller
range of kicks, however (Figure~\ref{amplitude2}). For velocity kicks
less than 350~km~s$^{-1}$, the amplitude--initial kick relation is:
$A(V_0)/{\rm kpc} = (8.06 \pm 0.51)\times 10^{-3}
V_0/\mathrm{km\:s^{-1}} + (0.126\pm 0.067)$, making the linear part of
the amplitude--kick relation independent of direction to within
$2\sigma$. Again, escape from the galaxy  occurs for an initial
kick $650$~km~s$^{-1}$.

\begin{figure}
\epsscale{.7}
\plotone{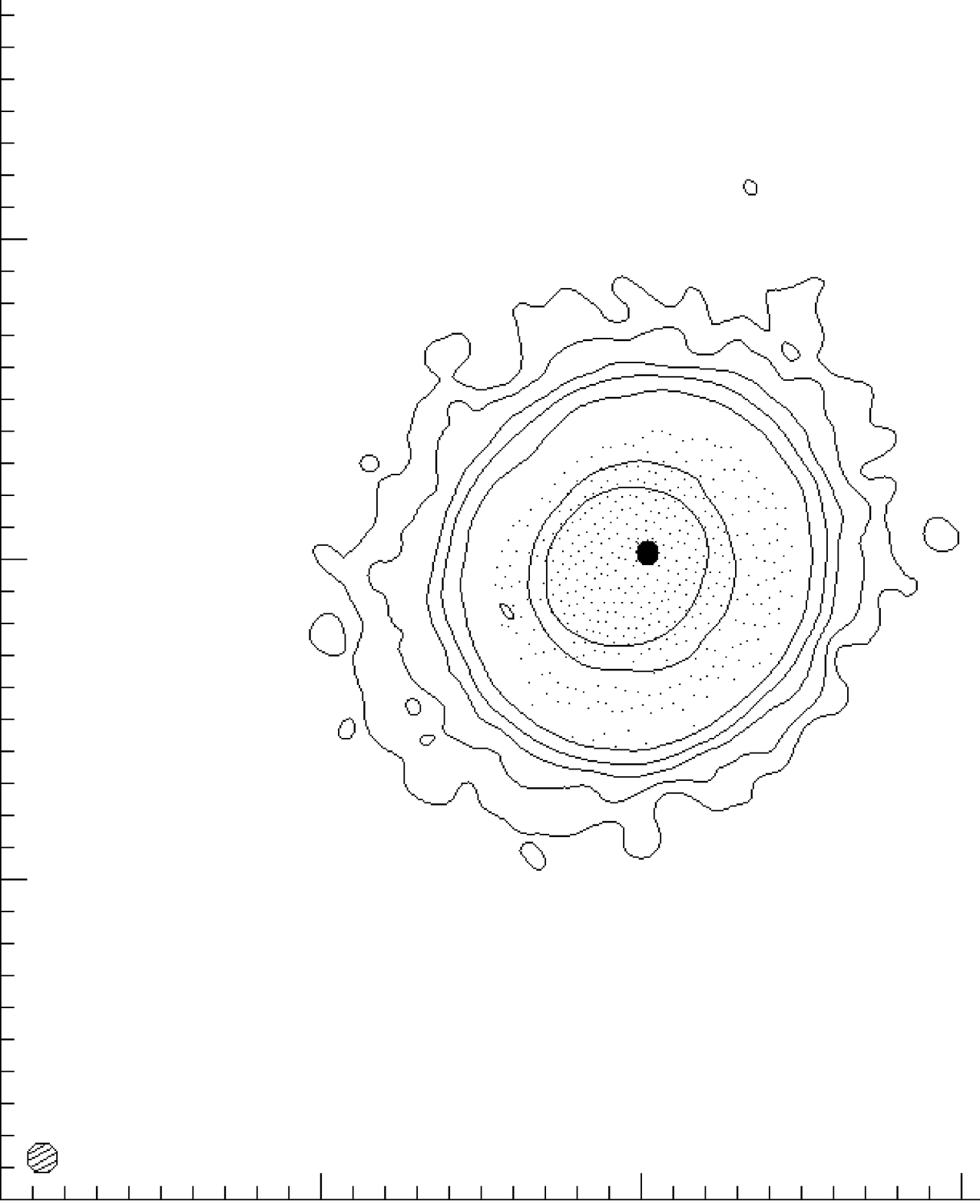}\\
\plotone{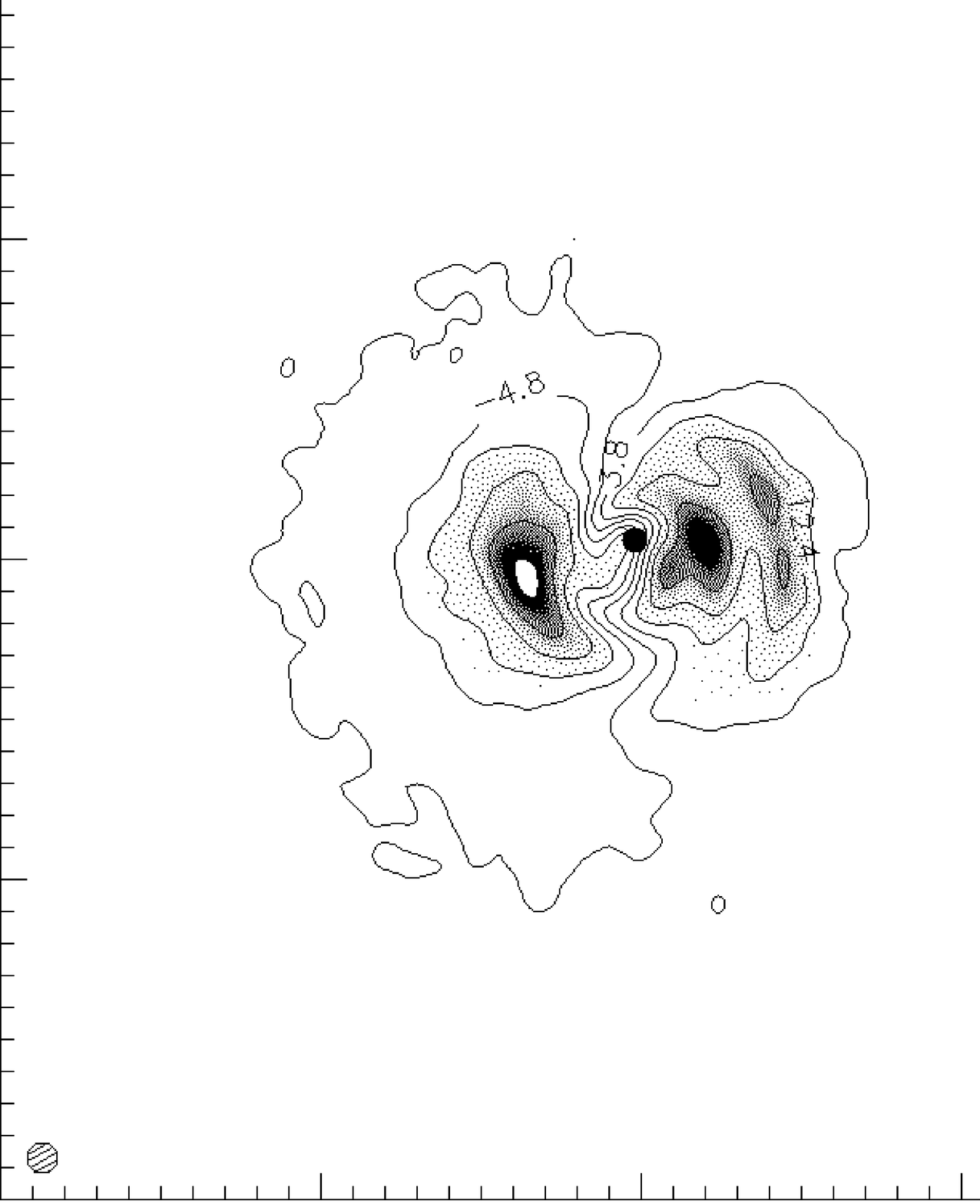}
\caption{Morphology and Kinematics of the galaxy model at $t=0.70$~Gyr
  after the central BH has been given a velocity kick of
  $300$~km~s$^{-1}$ perpendicular to the plane of the galaxy
  for a BH of mass $10^8~M_\odot$.  In the
  top panel, the stellar disk is illustrated in greyscale and the
  logarithmic contour lines indicate density of HI gas, after
  convolution with a simulated Gaussian beam (lower left). In the
  bottom panel,  the convolved radial velocity of the gas is indicated by the
  contour lines labeled in km~s$^{-1}$.
     The inclination to the line of sight is  $10$$^\circ$. In
  both panels the BH is indicated by the black dot in the center. Major
  tick marks delineate 25~kpc.}
\label{HI2}
\end{figure}

As in the coplanar case, little disturbance of the morphology and
kinematics of the stellar or gas disk is observed; only a displacement
of the BH from the center of the galaxy disk is observed. However,
when viewed nearly face--on, some streaming motions are observed in
the gas kinematics, as illustrated in Figure~\ref{HI2}.

\subsection{Black Hole Mass $10^9M_\odot$}
\begin{figure}
\epsscale{1}
\plotone{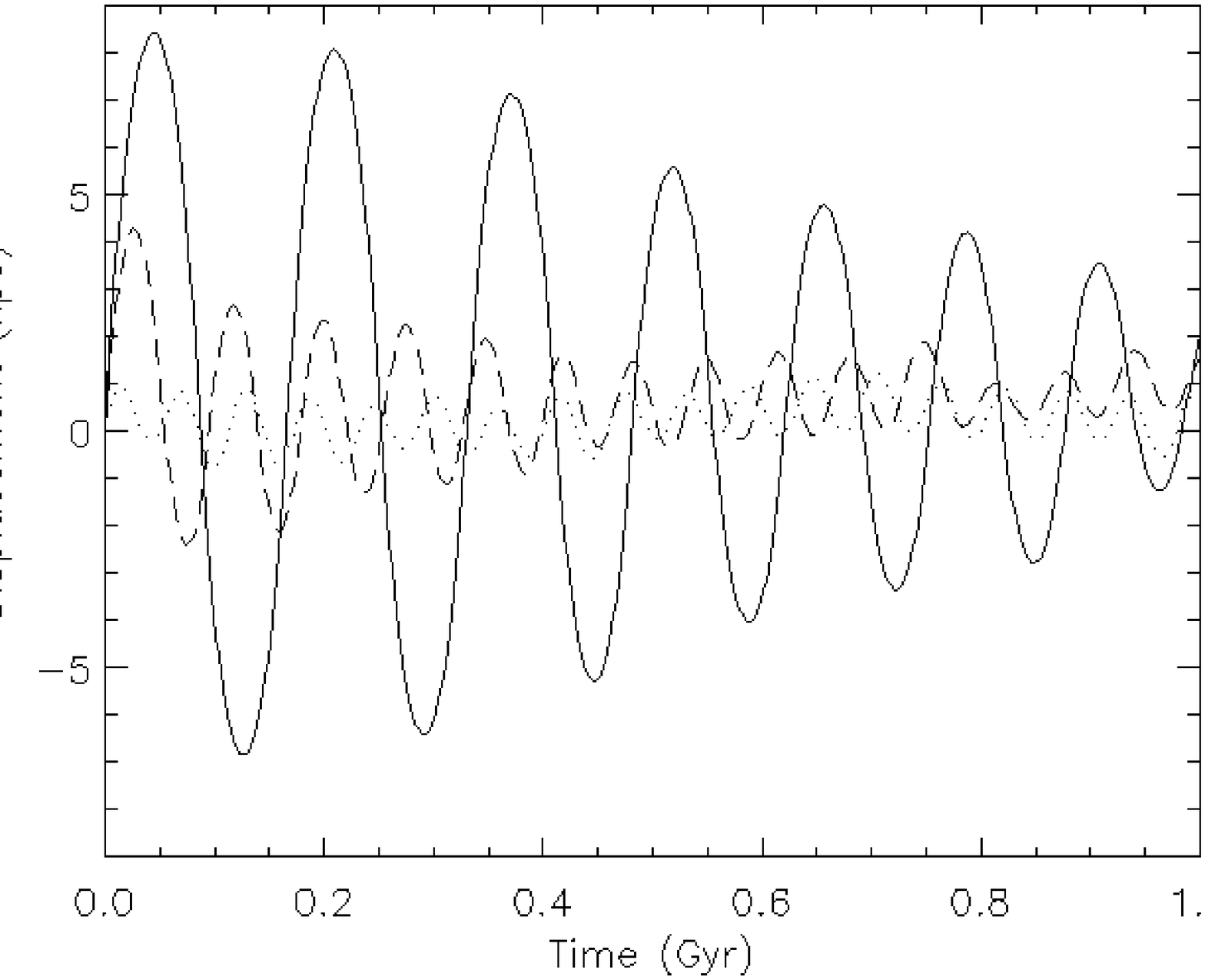}
\caption{Motion along the direction of kick, in the plane
  of the galaxy, for the Black hole of mass
  $10^{9}M_\odot$ inside a model disk galaxy for an initial velocity kick
  of $400$~km~s$^{-1}$ (solid line), $300$~km~s$^{-1}$ (dashed line), and
  $100$~km~s$^{-1}$ (dotted line). }
\label{XvsT2}
\end{figure}

When receiving an impulsive initial velocity ``kick,'' a large
$10^9M_\odot$ black hole exhibits qualitatively similar behavior to
the $10^8M_\odot$ black hole, as shown in Figure~\ref{XvsT2}. In this
case, however, it is clear that the damping time of the oscillation is
significantly shorter. In the $10^8M_\odot$ case, the damping scale
time was on the order of 1.5~Gyr; however, for the  $10^9M_\odot$
case, damping times are of the order of 0.8~Gyr.

During the period of significant motion of the black hole, significant
dynamical disturbance can be detected in the HI velocity maps. After a
damping scale time, however, the dynamics recover to a normal rotation
curve. Morphology outside the disk scale radius remains generally
undisturbed, although minor observable asymmetries occasionally
appeared in the central gas distribution. Despite the normal disk and
gas morphology, however, as in the previous case, the central black
hole could be found offset from the center of the disk by several kpc.

\begin{figure}
\epsscale{.7}
\plotone{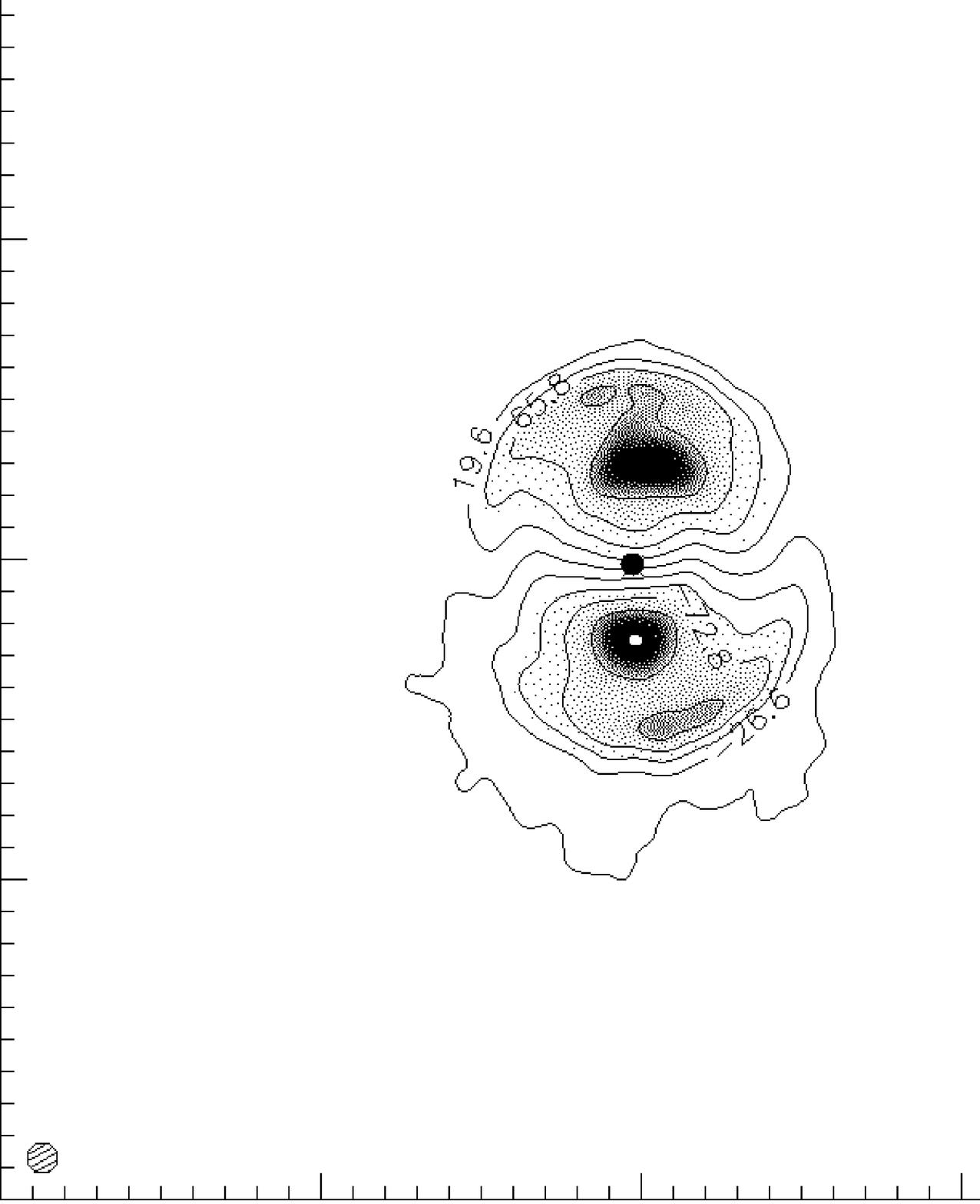}\\
\plotone{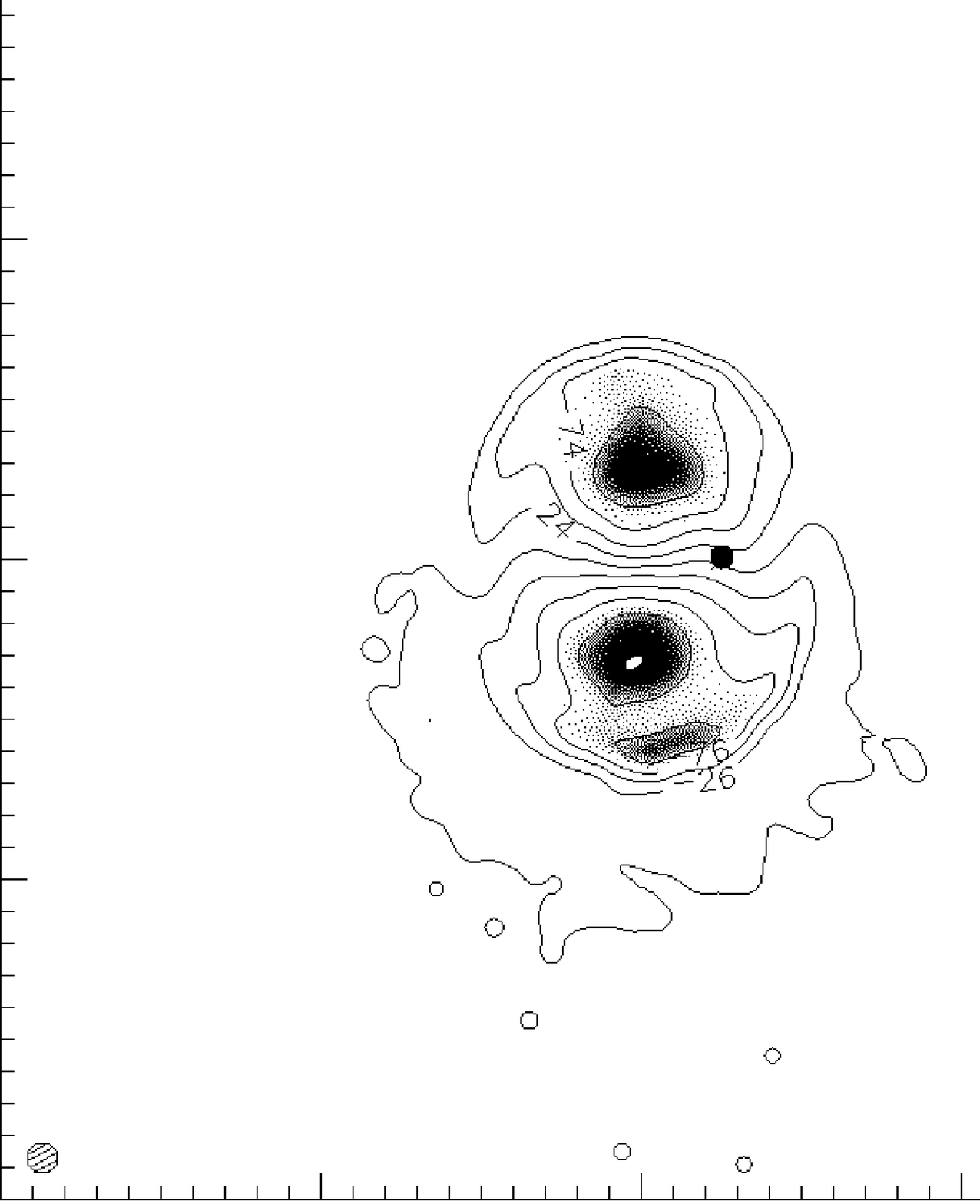}
\caption{Kinematic asymmetries in the gas component of a galaxy with a
 $10^9M_\odot$  BH  which has received an
  impulsive velocity kick. In the top panel, the kick velocity is
  $200$~km~s$^{-1}$ to the right in the plane of the galaxy.  The
  galaxy is inclined to the line of sight by
  $30$$^\circ$. 
      The  bottom panel is for the  same parameters but for
an initial velocity kick of $400$~km~s$^{-1}$.
   Contours of constant radial velocity are labeled in km~s$^{-1}$.  Major
  tick marks delineate 25~kpc.}
\label{VelPlots}
\end{figure}

Typical induced kinematic asymmetries are presented in
Figure~\ref{VelPlots}. In these plots, gas particle velocities were
convolved and combined in a simulated Gaussian telescope beam and
lines of constant radial velocity drawn. If the galaxy were a
symmetric rotator, then constant radial velocity lines should be
straight and parallel as they pass through the dynamical
center. However, as can be seen in the figure, these kinematics have
been disturbed by the motion of the BH, which has resulted in a
dynamical friction interaction with the galaxy. The central regions of
the galaxy, within the disk scale radius of $3.5$~kpc react quickly to
the transferred momentum of the BH, and tend to move with it in its
direction of motion. Beyond that radius, particles are shielded from
the gravity of the central BH and react sluggishly. As a result, the
dynamical center of the central galaxy is displaced from the overall
dynamical center in the direction of the initial impulse given to the
BH. In the transition region between the two regimes, streaming
motions are evident, with magnitudes on the order of
$30$~km~s$^{-1}$. Finally, because of the BH's resulting orbit around
the new dynamical center of the distribution, at most times its actual
location was offset from the dynamical center as in the second frame
of Figure~\ref{VelPlots}.

\section{BH Acceleration Due to a One-Sided Jet\label{BHAccel}}

    In order to simulate the influence of a one-sided jet, we did a
series of runs where the massive BH was placed in the center of the
galaxy with zero initial velocity relative to the galaxy, but with an
applied constant acceleration for a time of the order of the Salpeter
time.  The available accretion power from a BH is $L_a=\epsilon
\dot{M}_a c^2$ where $\dot{M}_a$ is the mass accretion rate and the
efficiency is assumed to be $\epsilon=0.1$.  Usually, the accretion
rate is measured in units of the rate which would give the Eddington
luminosity for the BH, that is $\dot{M}_{\rm Edd} = L_{\rm
Edd}/(\epsilon c^2) \approx 2.2(M_{bh}/10^8 M_\odot)
M_\odot$yr$^{-1}$, where $L_{\rm Edd} \approx 1.26\times
10^{46}(M_{bh}/10^8M_\odot)$ erg/s and $M_{bh}$ is the BH mass.  Thus,
the accretion luminosity can be written as $L_a = \dot{m} L_{\rm
Edd}$, where $\dot{m}\equiv \dot{M}_a/\dot{M}_{\rm Edd}$.  The
exponentiation time of the BH mass growth, the \citet{Salpeter} time,
is $T_S \approx 4.5\times 10^7 \dot{m}^{-1}$ yr.

     It is widely thought that the jets of active galaxies are due to
the presence of a strong, ordered magnetic field ($\sim 10^{3-4}$ G)
threading the BH accretion disk.  For the commonly considered case of
a field with dipole symmetry the magnetically driven jets are of equal
strength on the two sides of the disk \citep{Lovelace}.  However, as
\citet{Wang92} pointed out, an ordered field in the disk
consisting of {\it dipole} and {\it quadrupole} components can give
magnetically driven jets of unequal strength on the two sides of the
disk.  In fact, one side of the disk may have a dominant jet.  The
timescale for the jet to be on one side $\tau$ in this model is of the
order of the time for plasma and magnetic field to move through the
disk.  This time may be shorter than $T_S$.  Nevertheless, we
emphasize in our simulations the influence of a one-sided jet by
maintaining its force on the BH for a time of the order of $T_S$ after
which the force is zero.  We assume that the BH carries along with it
bound gas and stars sufficient to supply $\dot{M}_a$ during this time.

     For the case of a one-sided jet, we can write the jet luminosity
as $L_{\rm jet} = f_{\rm jet} L_a$, where $f_{\rm jet} \leq 1$ 
is the fraction of the accretion  luminosity going into the jet.  
Alternatively, for  asymmetrical,  oppositely
directed jets we take $L_{\rm jet} $ to be the
{\it difference} in the luminosities of the two jets.
  Assuming that the jet outflow is highly relativistic, the force on
the BH/disk system is therefore 
\begin{equation}
F_{bh}=-L_{\rm jet}/c \approx 4.2\times 10^{35} f_{\rm jet} \dot{m}
\left({M_{bh} \over 10^8 M_\odot}\right)~{\rm dynes}~,
\end{equation}
and the BH acceleration is
\begin{equation}
a_{bh} ={ | F_{bh}| \over M_{bh}}\approx 2.1\times 10^{-6}
 f_{\rm jet}\dot{m}~{{\rm cm}\over {\rm s}^2}~,
\end{equation}
\citep{Wang92}.   The momentum imparted to the
BH during a time $T$ is $I=\int_0^T dt ~M_{bh} a_{bh}
= T M_{bh} a_{bh}$ for constant one-sided jet.  
  For a jet which changes sides randomly on a timescale $\tau\ll T$,
the imparted momentum is  much smaller,
$I \sim  (\tau T)^{1/2}M_{bh} a_{bh}$.  
    The small values of the BH displacements estimated by
\citet{Wang92} resulted from an invalid assumption about
the rigidity of the central potential of the galaxy.   

   As a reference value for our simulations, we take 
$a_{bh} = 2\times 10^{-8}$ cm~s$^{-1}$ which corresponds to
$f_{\rm jet}\dot{m}=0.01$.  
  For this  value of $a_{bh}$, notice that
$\int_0^T dt ~a_{bh} = T a_{bh} \approx 
1260 (T/2\times10^8{\rm yr})$ km~s$^{-1}$.  
We confirmed the relevance of this value by observing that the hole
remains bound to the galaxy for applied 
accelerations up to $a_{bh}=4.5\times
10^{-8}$~cm~s$^{-2}$, and escapes the galaxy for larger $a_{bh}$.
    A hundred times larger value of $a_{bh}$
was considered by \citet{Tsygan07} to apply over a time scale
$T_S$ and this will strongly eject the BH from the galaxy.

\subsection{Black Hole Mass $10^{8}M_\odot$}
\begin{figure}
\epsscale{.8}
\plotone{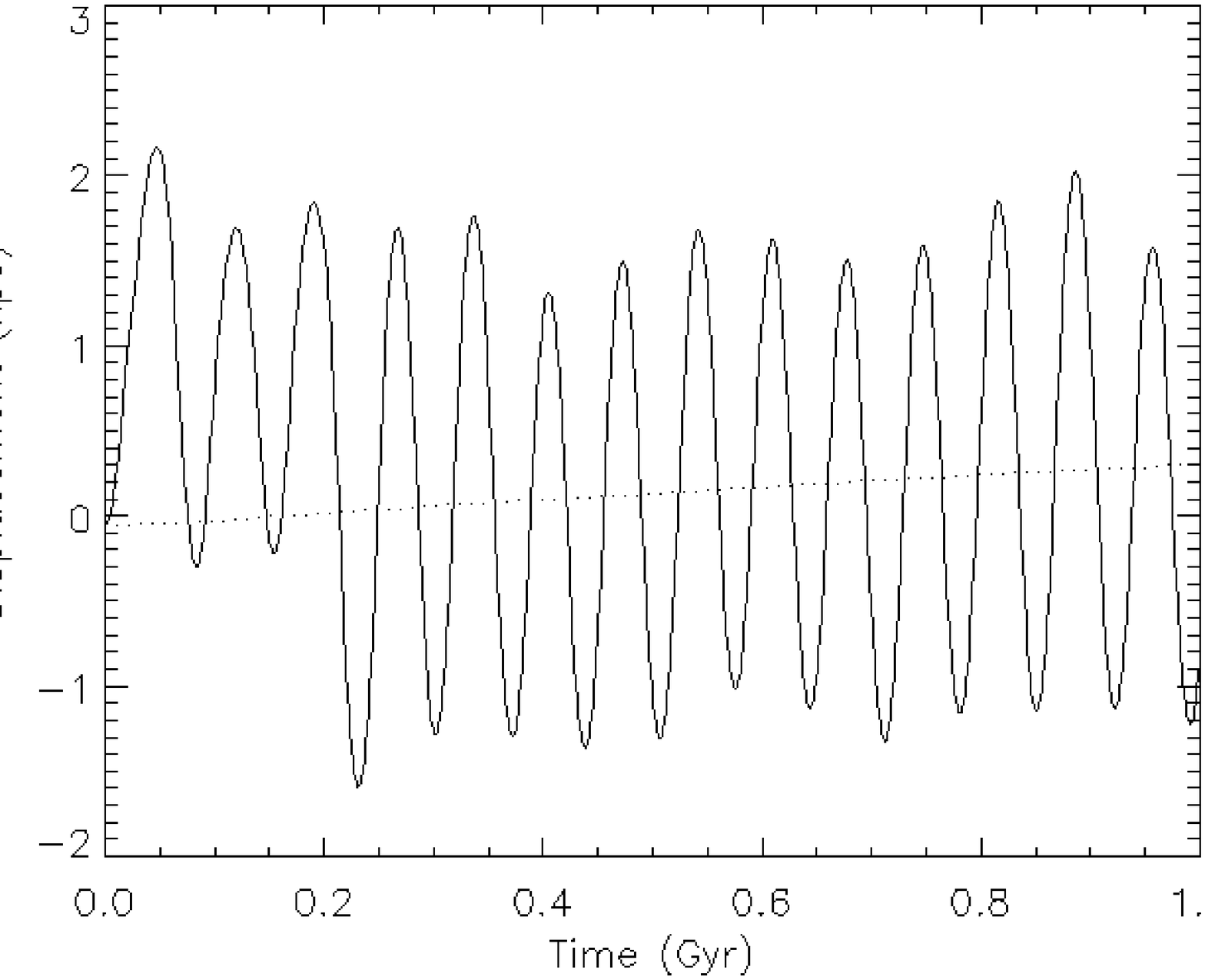}\\
\plotone{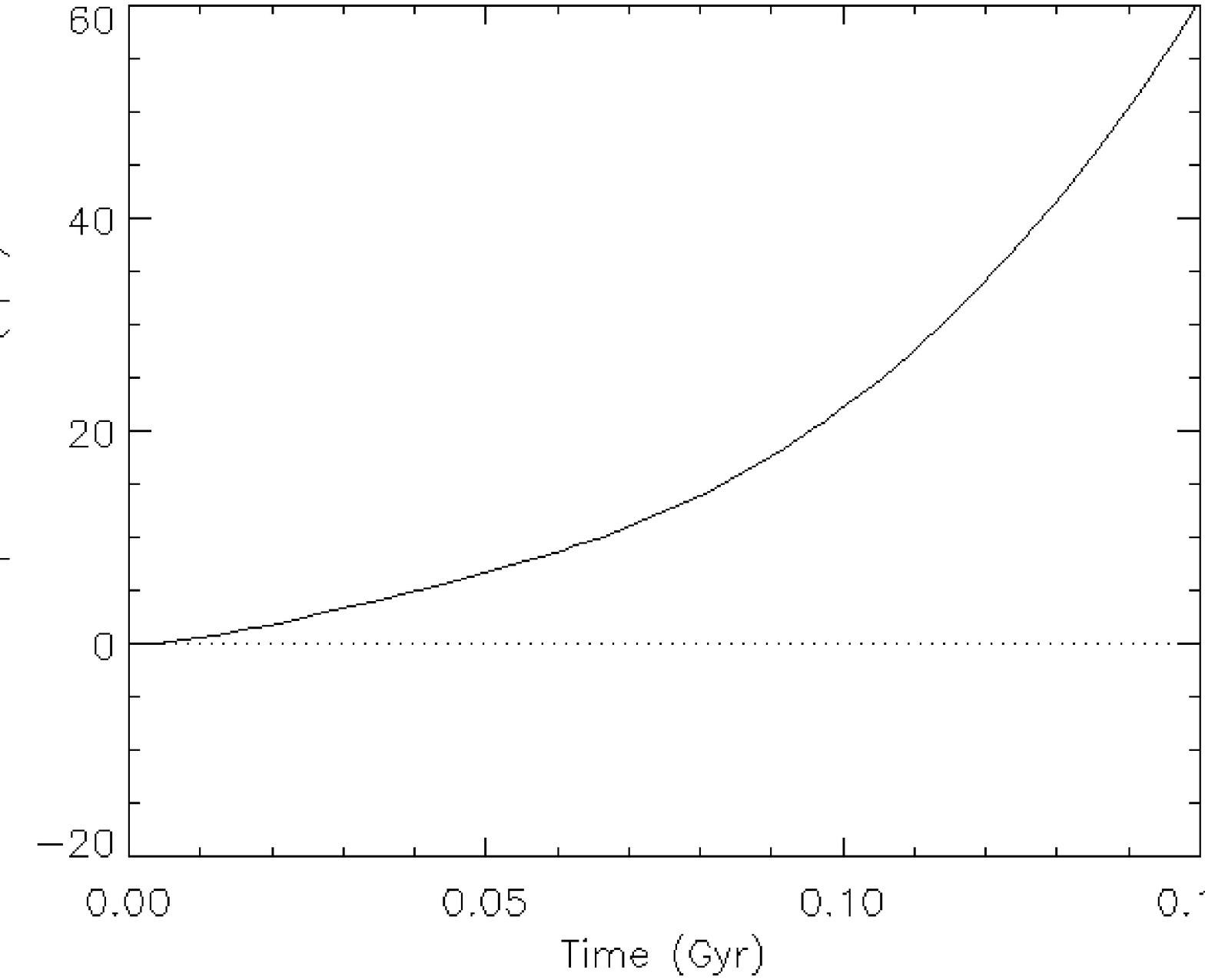}
\caption{Displacement vs. Time along the direction of applied
  acceleration in the plane of the galaxy for a $10^8M_\odot$ black
  hole with simulated jet acceleration of $2.0\times
  10^{-8}$~cm~s$^{-2}$ for a time $0\le t \le 0.2$~Gyr.  
   The dotted line shows the center
  of mass of the galaxy.   The bottom panel shows the same
  quantities but for an acceleration 
  $4.0\times 10^{-8}$~cm~s$^{-2}$.}
\label{XTdrag}
\end{figure}

When an external acceleration was applied in the plane of the galactic
disk, the typical behavior was for the $10^8M_\odot$ BH to find an
equilibrium position off--center of the morphological and dynamical
center of the galaxy and to oscillate about that position as
long as the acceleration was applied. Figure~\ref{XTdrag} illustrates
the BH displacement as a function of time for two typical examples of
this experiment, in which the $10^8M_\odot$ black hole was given an
applied acceleration of $2.0\times 10^{-8}$~cm~s$^{-1}$ and $4.0\times
10^{-8}$~cm~s$^{-1}$ in the plane of the galaxy. The center of mass of
the galaxy is also depicted for the first experiment. This accelerates
as well, as dynamical friction transfers momentum between BH and the
galaxy. In the second run, the BH escapes the galaxy entirely within
$0.1$~Gyr. In each run where the BH remained bound, we confirmed that
the acceleration of the center of mass of the galaxy was equal to
$a_{bh}(M_{bh}/M_{\rm tot})$, where $a_{bh}$ was the acceleration
imparted to the BH, and also that when $a_{bh}$ was turned off after
$200$~Myr, that the final velocity of the center of mass agreed with
the total impulse $I=\int_0^{T_S}dt M_{bh}a_{bh}$ given, within errors
which were taken to be the observed fluctuation in the center of mass
velocity and acceleration of a galaxy with no kick (approximately
$10\%$).

\begin{figure}
\epsscale{.7}
\plotone{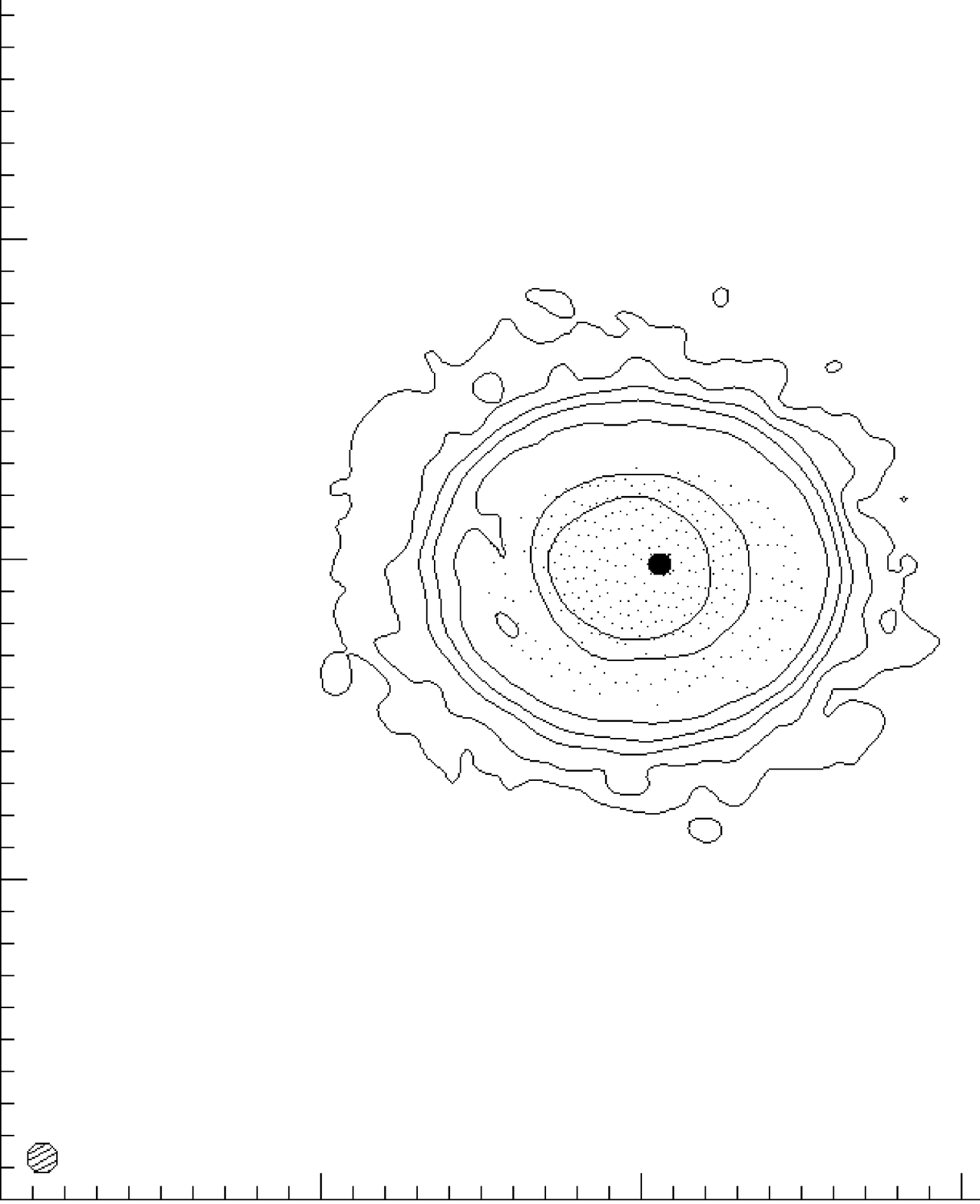}\\
\plotone{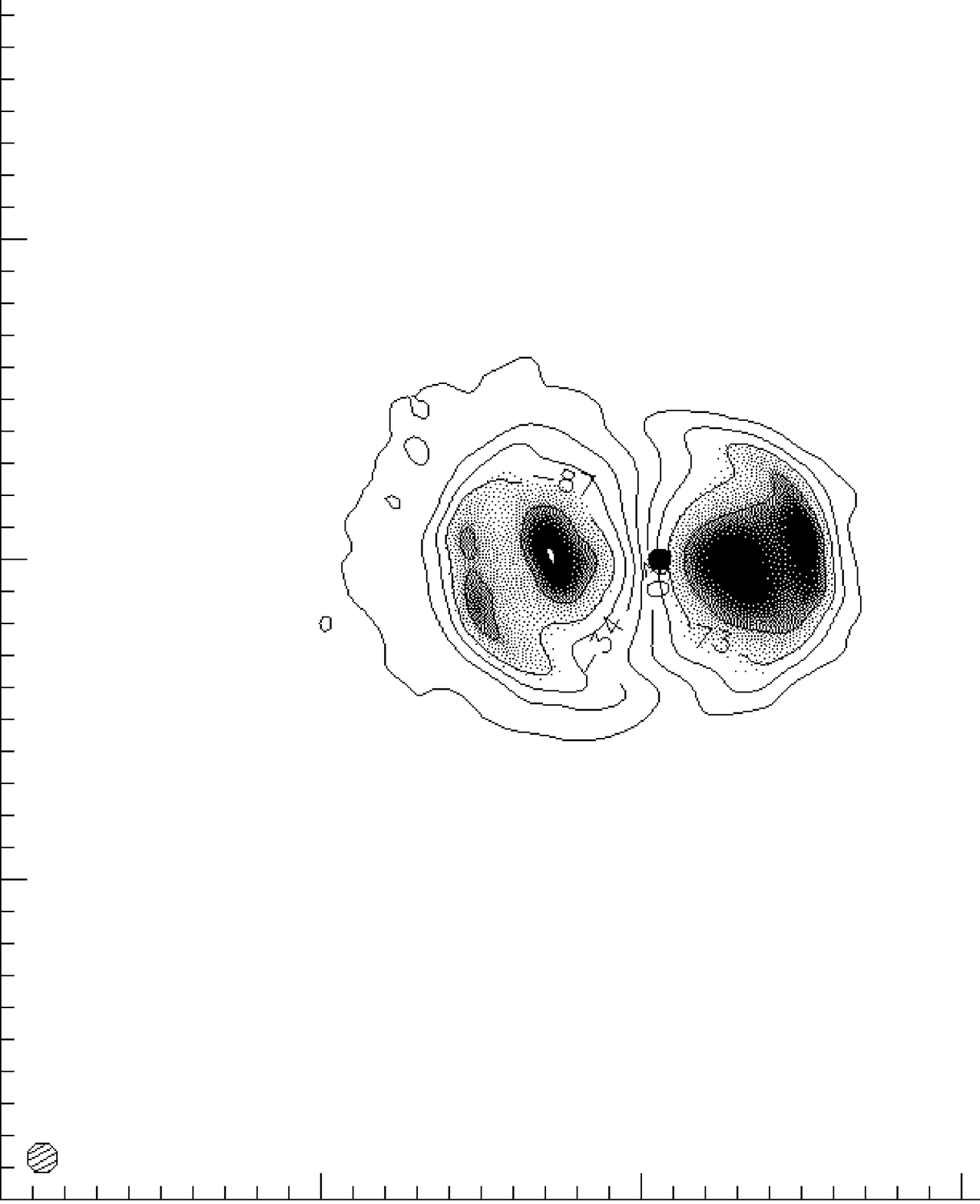}
\caption{Galaxy model gas density contours and velocity map, convolved
  with a simulated Gaussian telescope beam, for the $10^8M_\odot$
  BH with an applied acceleration of $2.0\times
  10^{-8}$~cm~s$^{-2}$ at time $t=0.2$~Gyr. 
      The inclination to line of
  sight is 30$^\circ$ about the axis of acceleration.  Major
  tick marks delineate 25~kpc.}
\label{CV1}
\end{figure}

For BHs of mass $10^8M_\odot$, slight to moderate asymmetries
were induced in the dynamics of the host galaxy during the ``on''
phase, although nothing detectable was induced in the morphology, as
illustrated in Figure~\ref{CV1}. When viewed with the galaxy inclined
to the line of sight, with the axis inclination aligned with the
applied acceleration, deformations of the contours of constant radial
velocity of order $40$~km~s$^{-1}$ are clearly observed in the
kinematics of the galaxy gas. Note that no grand--design spiral arms
have formed in the stellar or gaseous components, ruling out density
waves as a possible explanation for these. These streaming motions are
also present when the line of sight is at other orientations to the
line of sight, but are less prominent. These disk asymmetries
disappeared when the acceleration was turned off, but the BH
continued to oscillate about the center of mass with an amplitude that
depended slightly on the initial acceleration, but was typically of
order $1-2$~kpc.

\begin{figure}
\epsscale{1}
\plotone{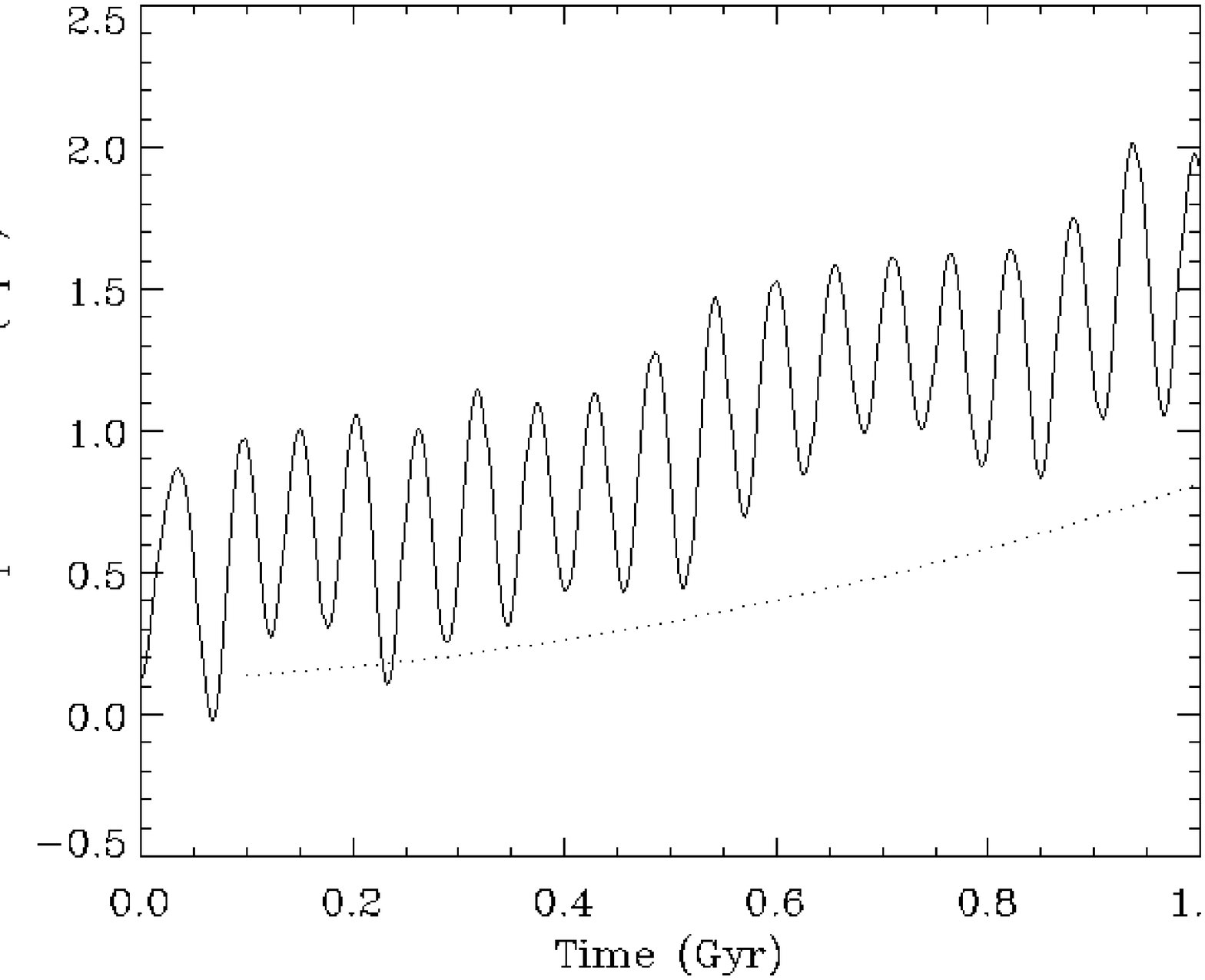}
\caption{The BH displacement as a function 
of time along the direction of the applied
  acceleration perpendicular to the plane of the galaxy for a $10^8M_\odot$ black hole.
      The  simulated jet acceleration is $2.0\times
  10^{-8}$~cm~s$^{-1}$. 
 The dotted line shows the center of mass of the galaxy.  }
\label{XTydrag}
\end{figure}

In the case where the BH is given an acceleration
perpendicular to the galaxy plane, a similar behavior is observed, as
illustrated in Figure~\ref{XTydrag}. The BH oscillates around
an equilibrium point, slowly dragging the galaxy along with it via
dynamical friction. In this case, however, the
amplitude of the motion is smaller and the frequency higher. The
acceleration required for escape in this case is $3.0\times
10^{-8}$~cm~s$^{-2}$, as compared with $4.5$ for the case in the plane
of the galaxy. This case produces no detectably significant
asymmetries in the morphology or kinematics of the galaxy disk.

\subsection{Black Hole Mass $10^9M_\odot$}
\begin{figure}
\epsscale{.8}
\plotone{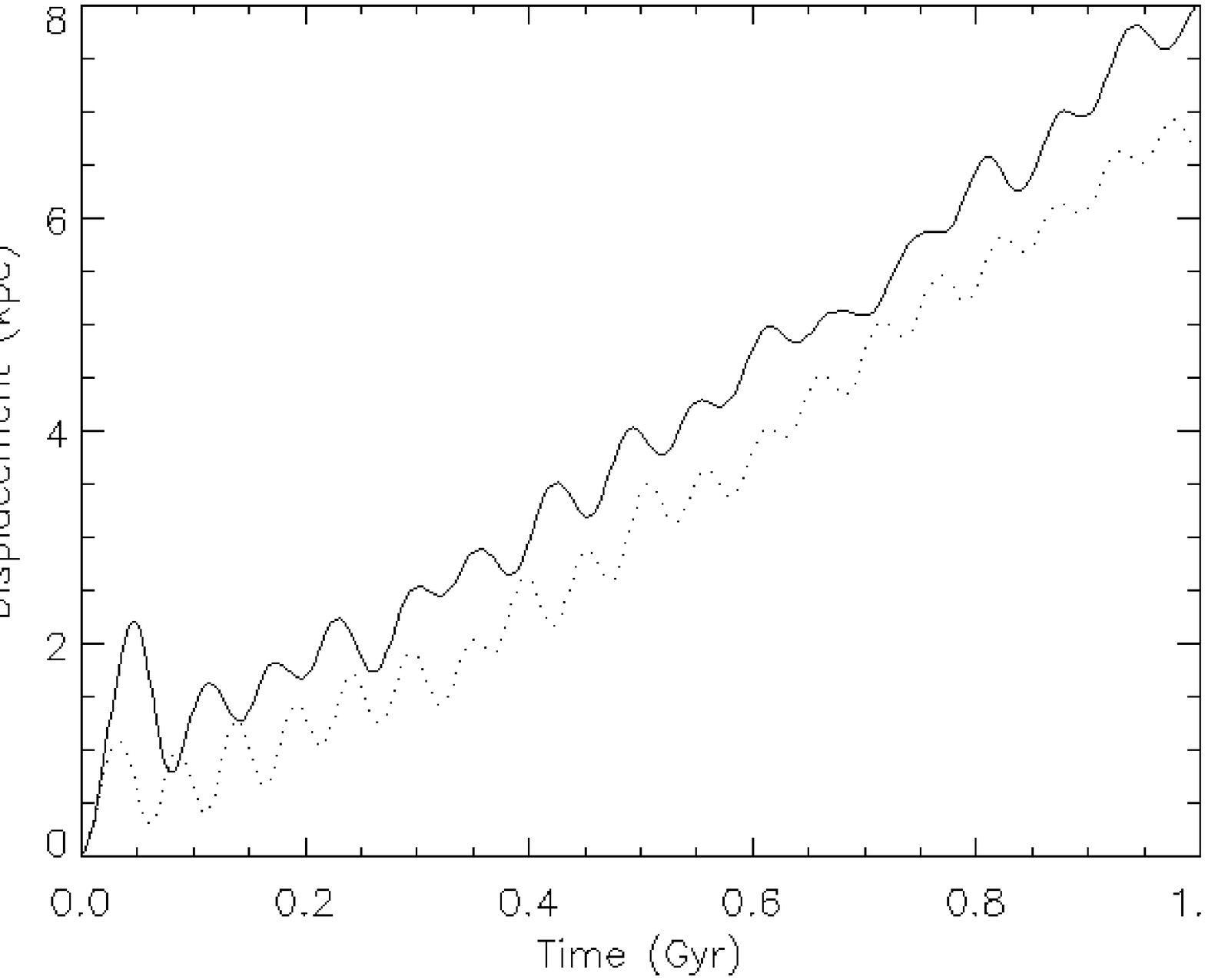}\\
\plotone{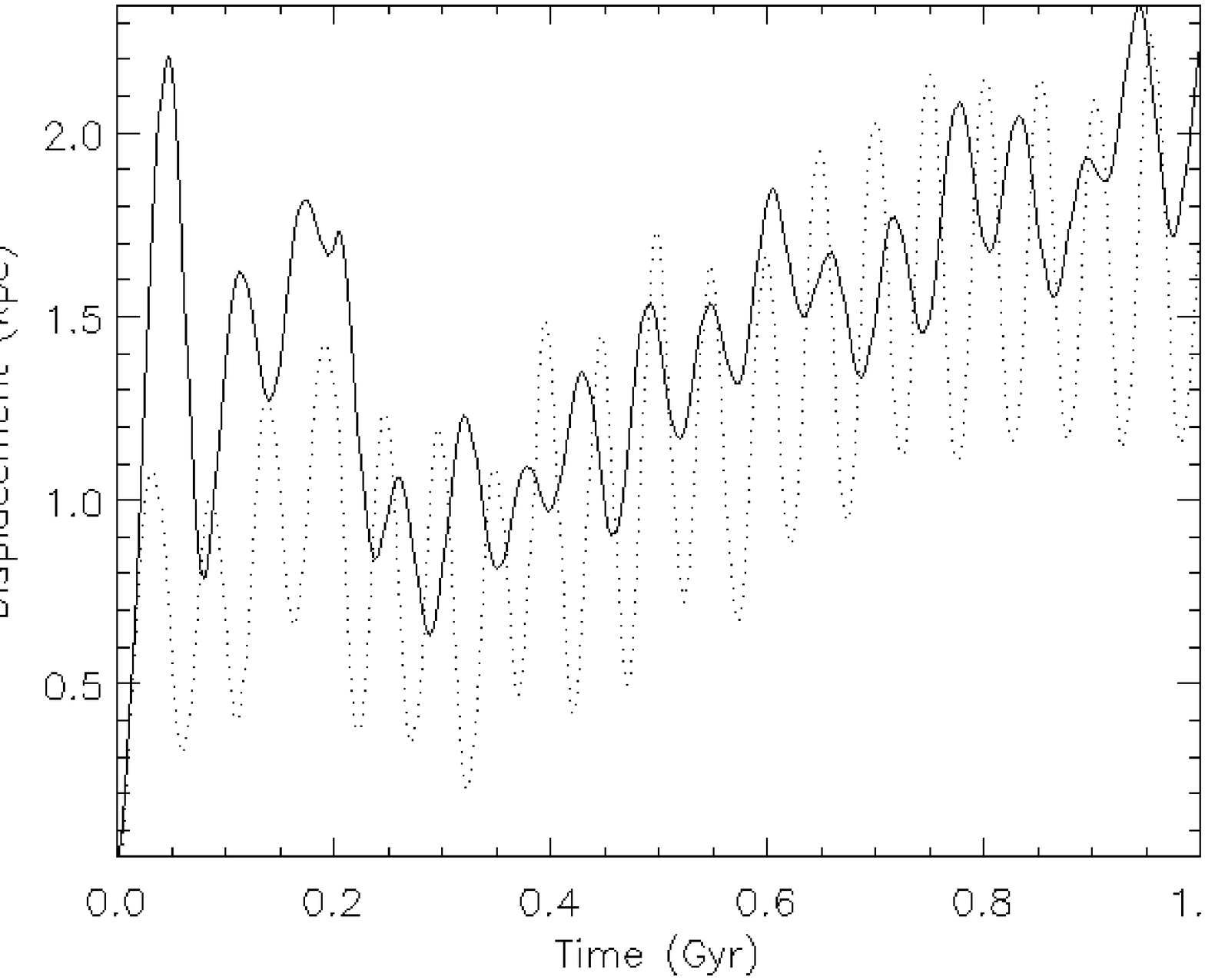}
\caption{Displacement vs. Time along the direction of applied
  acceleration for the black hole of mass $10^9M_\odot$ with
  simulated jet acceleration of $2.0\times 10^{-8}$~cm~s$^{-2}$. 
  Acceleration in the plane of the galaxy is represented by the solid
  line; perpendicular to the plane by the dotted line. In the bottom
  panel, the acceleration has been shut off at $0.2$~Gyr.}
\label{XT109}
\end{figure}

When given an acceleration in the plane of the galactic disk, the
large $10^9M_\odot$ black hole exhibited behavior similar to that of
the smaller hole.  In this case, as illustrated in Figure~\ref{XT109},
the larger impulse ``drags'' the galaxy a proportionally greater
distance. When the acceleration is turned off after $0.2$~Gyr, the
velocity imparted to the galaxy is $200$~m~$s^{-1}$.

\begin{figure}
\epsscale{.7}
\plotone{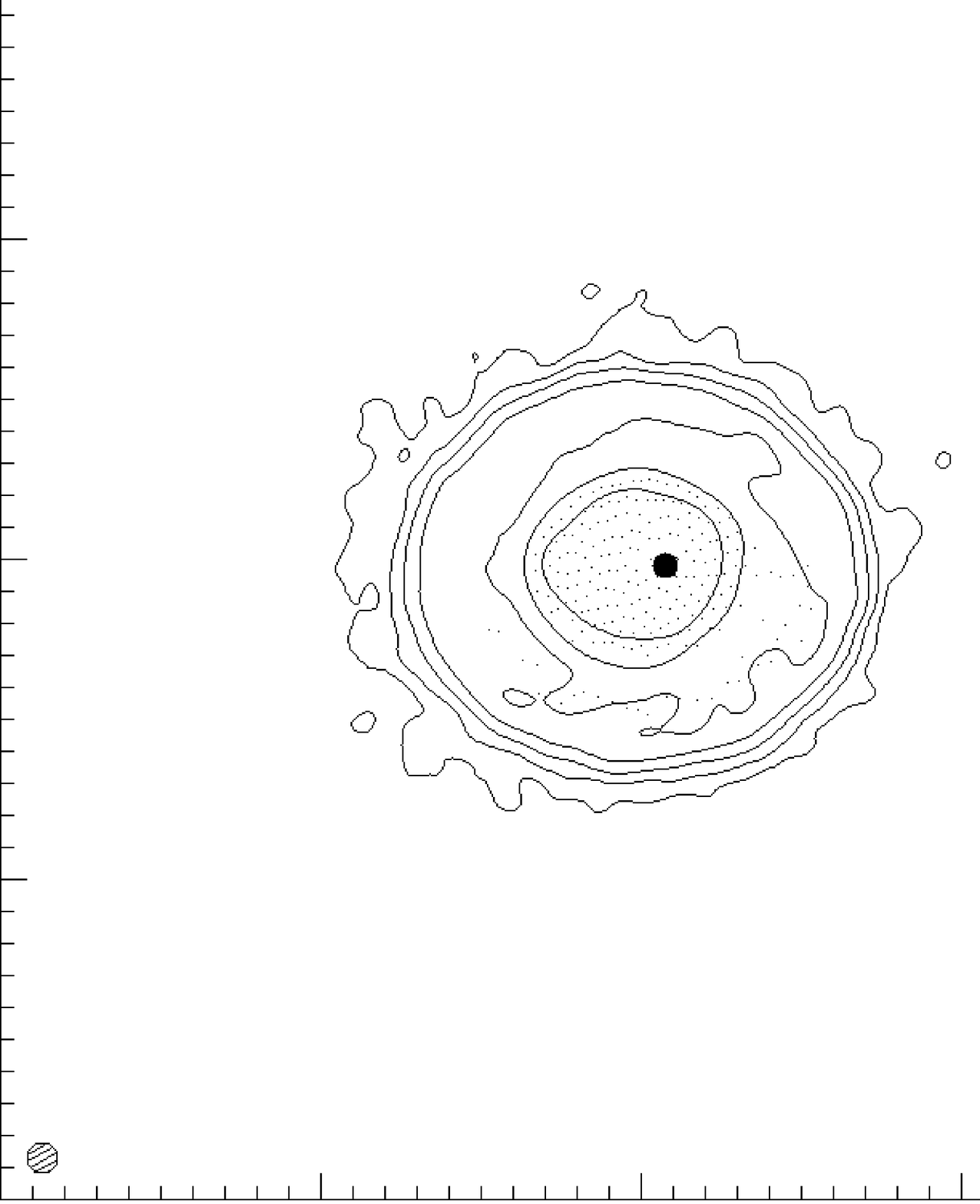}\\
\plotone{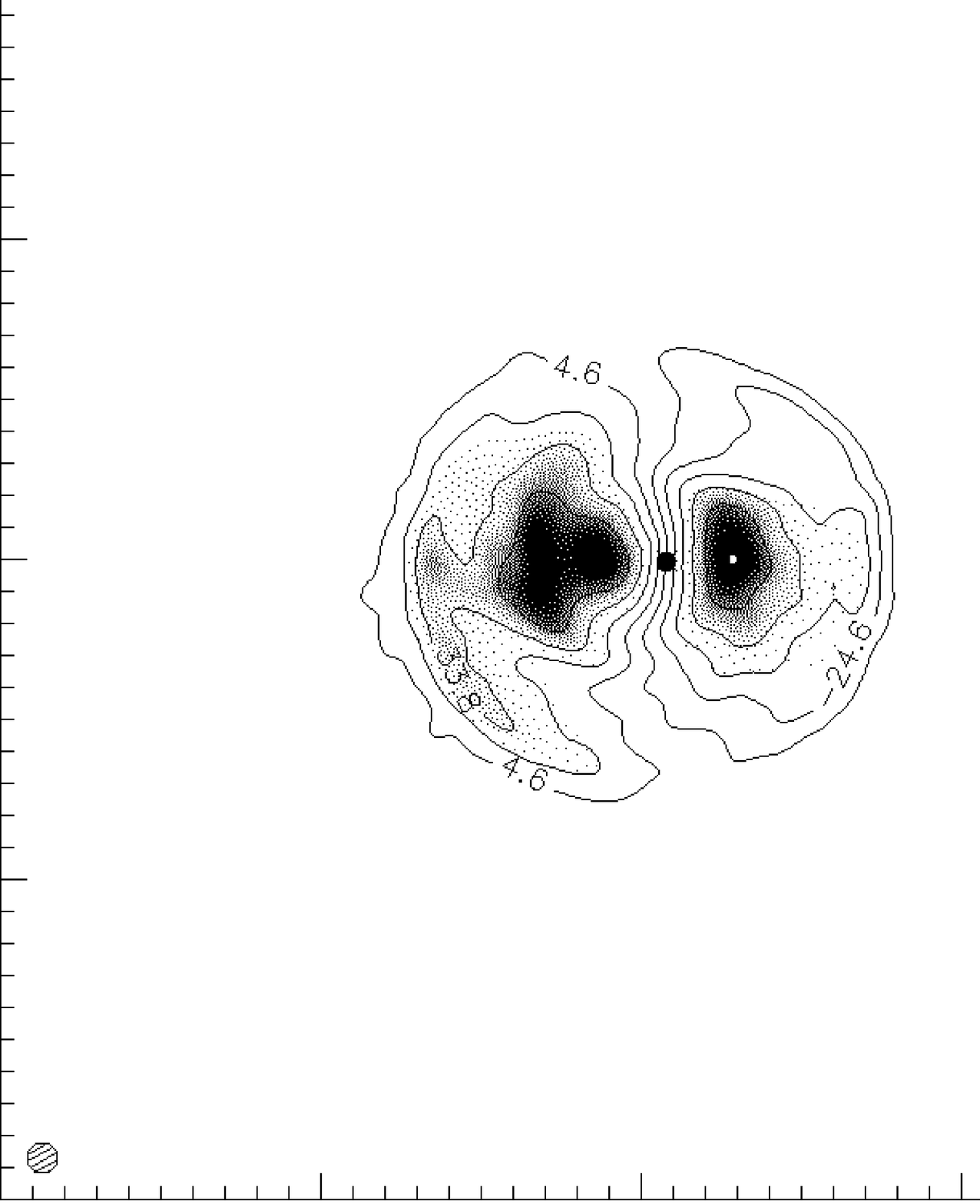}
\caption{Galaxy model gas density contours (top panel) and velocity
  map (bottom panel), convolved with a simulated Gaussian telescope
  beam, for the case of a $10^9M_\odot$ black hole with an applied
  acceleration of $2.0\times 10^{-8}$~cm~s$^{-2}$ in the plane of the
  galaxy, at time $t=0.2$~Gyr.  The inclination to line of sight is
  $30$$^\circ$ about the axis of acceleration. Major
  tick marks delineate 25~kpc.}
\label{CV2}
\end{figure}

As the black hole was accelerated, it became displaced from the
morphological and dynamical center of the galaxy, as illustrated in
Figure~\ref{CV2}. These induced asymmetries in the galaxy disk are
more pronounced in this case than in the $10^8M_\odot$ case, in
particular tending to drag the central regions of the galaxy with the
movement of the black hole via dynamical friction. In particular,
while the BH in the first frame of Figure~\ref{CV2} appears displaced
from the morphological center of the galaxy, the second frame of the
figure illustrates how the dynamical center of the galaxy has been
displaced towards the direction of acceleration of the black hole. The
result is a characteristic ``tongue--'' shaped extension of the
velocity contours on the side of the galaxy opposite the acceleration
and flattened contours on the side of the galaxy in the direction of
the acceleration. In test runs with a much larger, possibly unphysical
$10^{10}M_\odot$ BH, these features became associated with a central
core region of the disk that moved along with the black hole, being
more coupled to it than to the galaxy, and causing strong
morphological asymmetries in the disk as the bulge was dragged
off--center. We did not observe such strong asymmetric features in the
morphology of the $10^9M_\odot$ galaxy, but do observe the ``tongue''
features that indicate that the BH is beginning to dominate the
dynamics of the inner $2-5$~kpc of the disk.

\begin{figure}
\epsscale{.7}
\plotone{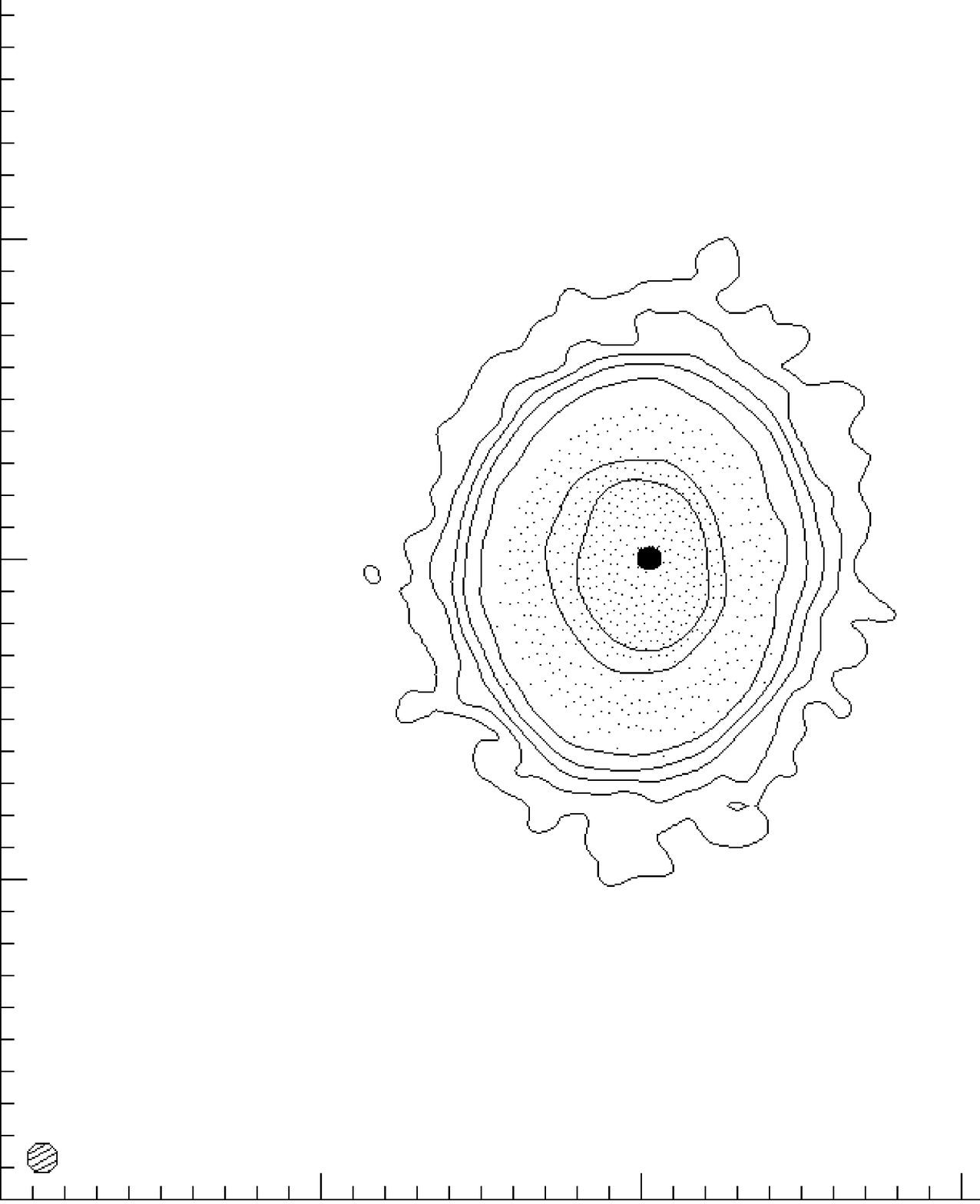}\\
\plotone{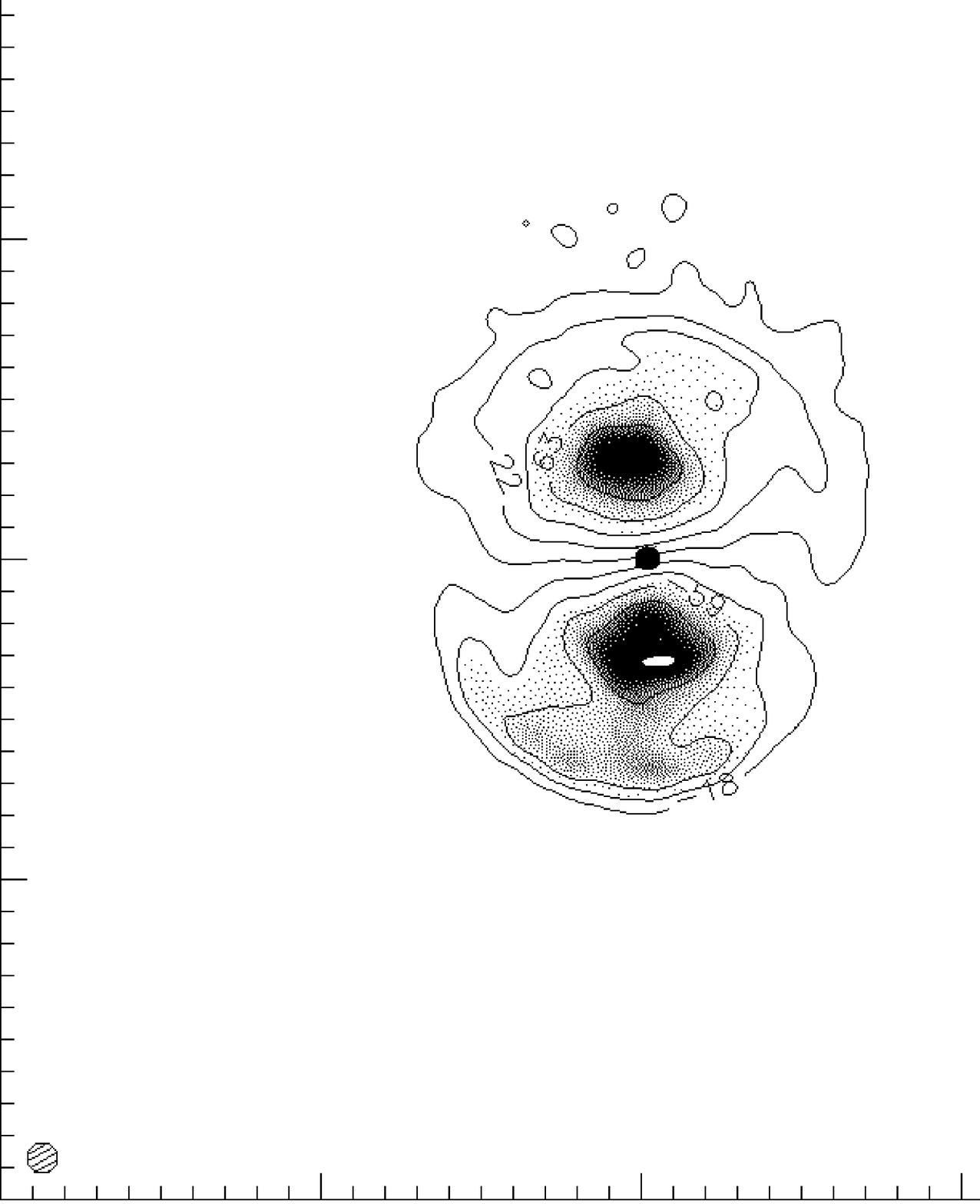}
\caption{Gas density contours (top panel) and velocity map (bottom
  panel), convolved with a simulated Gaussian telescope beam, for the
  case of a $10^9M_\odot$ black hole with an applied acceleration of
  $2.0\times 10^{-8}$~cm~s$^{-1}$, perpendicular to the plane of the
  galaxy at time $t=0.2$~Gyr.  The inclination to the line of sight is
  $30$$^\circ$. Major
  tick marks delineate 25~kpc. }
\label{CV101}
\end{figure}

A similar behavior is observed when the acceleration is perpendicular
to the plane of the galaxy.  Again, streaming motions and dynamical
asymmetries are observed for the first $0.5$~Gyr of acceleration, as
shown in Figure~\ref{CV101}.  Following that period, only slight
streaming motions remain as part of the stabilized disk kinematics.

\section{Discussion\label{Results}}

We have analyzed the motions of massive black holes in the centers of
galaxies which have received an impulsive kick (owing to the merging
of two two BHs) or received a sustained acceleration due to a one
sided jet.  We have studied the BH displacement from the galaxy center
and the influence this has on the galaxy morphology and kinematics.
An impulsive ``kick'' exceeding 600~km~s$^{-1}$ will ejects the BH
from the galaxy.  Accelerations of the order of $4\times
10^{-8}$~cm~s$^{-2}$ or larger for times of the order of $2\times
10^8$ yr eject the BH from the galaxy.  This acceleration is smaller
than the Eddington limit on the one-sided jet luminosity \citep{Tsygan07}
by a factor of about $50$.

For smaller velocity kicks or smaller sustained accelerations, the BH
is observed to oscillate through the galaxy center for times of order
Gyr.  The frequency of these oscillations decreases as the BH velocity
kick increases or as the BH acceleration increases.  In such galaxies,
we can expect to observe central BHs which are offset from the centers
of their host galaxy disks or have large radial velocity differences
from their hosts. Larger mass BHs undergoing these oscillations induce
streaming motions in their galaxies due to dynamical friction. This
process may be a source of disk morphological and dynamical asymmetry
in galaxies such as NGC~1637 or NGC~991 which exhibit nonaxisymmetries
but are located in isolated fields far from potential sources of tidal
interactions \citep{K98}.

The asymmetries in the galaxy disk induced by the BH motion are of
course more pronounced for more massive black holes.  The BH tends to
drag the central regions of the galaxy with it owing to dynamical
friction.  The dynamical center of the galaxy is seen some cases to be
displaced towards the direction of the BH acceleration with a
characteristic ``tongue--'' shaped extension of the velocity contours
on the side of the galaxy opposite the acceleration.

As a check on the results, additional model runs with fewer $(10^5)$
or more $(10^7)$ particles yielded substantially similar behavior for
the BH. Asymmetries induced in the galaxy disk remained qualitatively
similar for each run, although the details of the locations and
intensities of the induced asymmetries varied somewhat.

It is important to extend these studies to elliptical galaxies in the
future. We expect that the results concerning conditions of BH escape
from the galaxy will apply equally well there, as escape times from
the disk are very short, of order $0.1$~Gyr, and after that the
gravitational potential is essentially that of the spheroidal halo.

\acknowledgments
We thank Martha Haynes and Larry Kidder for
discussions. This work has made use of the computational facilities of
the National Astronomy and Ionosphere Center, which is operated by
Cornell University under a cooperative agreement with the National
Science Foundation.

\end{document}